\documentclass[%
 reprint,
 superscriptaddress,
showpacs,preprintnumbers,
 amsmath,amssymb,
 aps,
 prb,
 longbibliography,
]{revtex4-1}

\usepackage{graphicx}
\usepackage{dcolumn}
\usepackage{bm}
\usepackage{color}
\usepackage{ulem}

\begin{document}

\preprint{APS/123-QED}

\title{
Magnetism in S=1/2 Double-Perovskites with Strong Spin-Orbit Interactions
}

\author{Hiroaki Ishizuka}
\affiliation{
Kavli Institute for Theoretical Physics, University of California, Santa Barbara, California 93106, USA
}

\author{Leon Balents}
\affiliation{
Kavli Institute for Theoretical Physics, University of California, Santa Barbara, California 93106, USA
}

\date{\today}

\begin{abstract}
We study magnetism on the fcc lattice with tetragonal distortions, with general exotic directional magnetic interactions allowed by symmetry.  We consider two models, corresponding to a uniform tetragonal distortion, or a two-sublattice model with a tetragonal screw axis.  We establish their low temperature phase diagrams in the semi-classical limit using classical optimization and consideration of fluctuations both analytically and by Monte Carlo simulation. Both order by disorder and exchange anisotropy mechanisms favor a $\langle 110\rangle$ easy axis for magnetization.  We also show that spin-lattice coupling can give rise to an intermediate temperature paramagnetc nematic/orthorhombic phase, and discuss the transitions to/from this state.  These results are relevant to a family of insulating magnetic double perovskites, and find immediate application to the ferromagnet Ba$_2$NaOsO$_6$.  
\end{abstract}

\maketitle

\section{Introduction}

Spin-1/2 antiferromagnets are of considerable intrinsic interest because of their potential for strong quantum effects.  Of special value conceptually are ``pure'' spin systems without orbital degeneracy, where the magnetism can be isolated from other phenomena such as the Jahn-Teller effect and Kugel-Khomskii exchange.  Strong spin-orbit coupling (SOC), as is present in heavy 5d transition metal ions, provides a novel means of ``purifying'' orbitally degenerate spins by spin-orbital entanglement.  This mechanism leads to novel directional-dependent exchange coupling of spins, as has discussed in many iridates~\cite{Cao2003,Singh2010,Liu2011,Choi2012,OMalley2008,Singh2012,Modic2014,Takayama2014,Biffin2014} and a wide family of insulating double perovskites.  In this paper, we discuss a specific example of the latter, in which S=1/2 spins interact via novel anisotropic interactions on the geometrically frustrated face-centered cubic (fcc) lattice.  We find a rich structure of magnetic phases, examples of thermal and quantum order-by-disorder, and the possibility for an intermediate nematic phase induced by spin-lattice coupling. 

With strong SOC and ideal cubic symmetry, spins and orbitals in double perovskites with a single 5d electron per transition metal site combine to form an effective $S=3/2$ spin residing on a face-centered cubic (fcc) lattice.  This ``large'' spin may still be rather quantum due to unusual multipolar interactions, but may also be reduced to an effective $S=1/2$ one by a structural transition or quadrupolar ordering, either of which may reduce the cubic symmetry.  Prior theoretical work in an $S=3/2$ model suggested a transition to tetragonal symmetry indeed may occur.  If this transition occurs at a temperature large compared to magnetic scales, an $S=1/2$ Hamiltonian can provide a more economical description of the latter.  Such a cubic to tetragonal transition has indeed been observed recently in the material Ba$_2$NaOsO$_6$ at 320K.~\cite{IslamUP}  Since the magnetic order sets in only around 10K in this material, it is a strong candidate for the $S=1/2$ description.

With the general problem and this specific material as motiviation, we study here pure $S=1/2$ spins on a tetragonally distorted fcc lattice.  We consider two models.  In the first, the tetragonal state is produced by a simple expansion or contraction of the cubic $z$ axis.  In the second, ``twisted'' model, motivated by the theory of Ref.~\onlinecite{Chen2010}, we consider a different tetragonal state with an doubled unit cell along the $z$ axis.  We study these models by a combination of classical analysis, thermal and quantum spin wave theory, and Monte Carlo simulation.  We derive in this way phase diagrams over a broad parameter space, which should be applicable to a wide range of double perovskite materials.  

\begin{figure}
   \includegraphics[width=\linewidth]{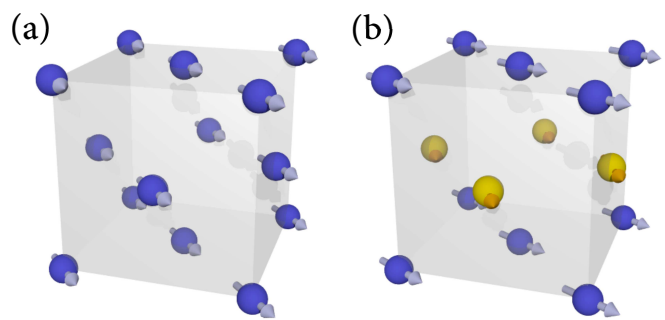}
   \caption{
   Magnetic structures on the fcc lattice: (a) ferromagnetic state with $\langle110\rangle$ anisotropy and (b) canted ferromagnetic phase. The spheres denote Os sites and the different colors in (b) indicate different ionic environments. See Sec.~\ref{sec:stagg-tetr-case} for details. 
   }
   \label{fig:intro:fcc}
\end{figure}

An interesting feature of the simple tetragonal model is a mean-field U(1) degeneracy of a set of ferromagnetic and antiferromagnetic states with spins ordered within the XY planes.  In these states, spins may be rotated by an arbitrary global angle with no energy cost.  This is an accidental degeneracy which is not required by the tetragonal $C_4$ symmetry of the model.  Hence one expects, and we indeed find, that the symmetry is broken by fluctuations, both classical thermal and zero point quantum ones.  Remarkably, fully independently of the Hamiltonian parameters, these fluctuations select the four states with magnetization oriented along the $\langle 110\rangle$ axes [See Fig.~\ref{fig:intro:fcc}(a)].  This selection is in fact precisely what has been observed experimentally in Ba$_2$NaOsO$_6$ as the ferromagnetic easy axis.  We also show that the same $\langle 110\rangle$ axes are preferred in the twisted model, though in this case the selection occurs already at the classical level.  In this case, the low temperature phase is a canted state with simultaneous and orthogonal staggered and uniform moments within the XY plane [See Fig.~\ref{fig:intro:fcc}(b)].

\begin{figure}
   \includegraphics[width=0.85\linewidth]{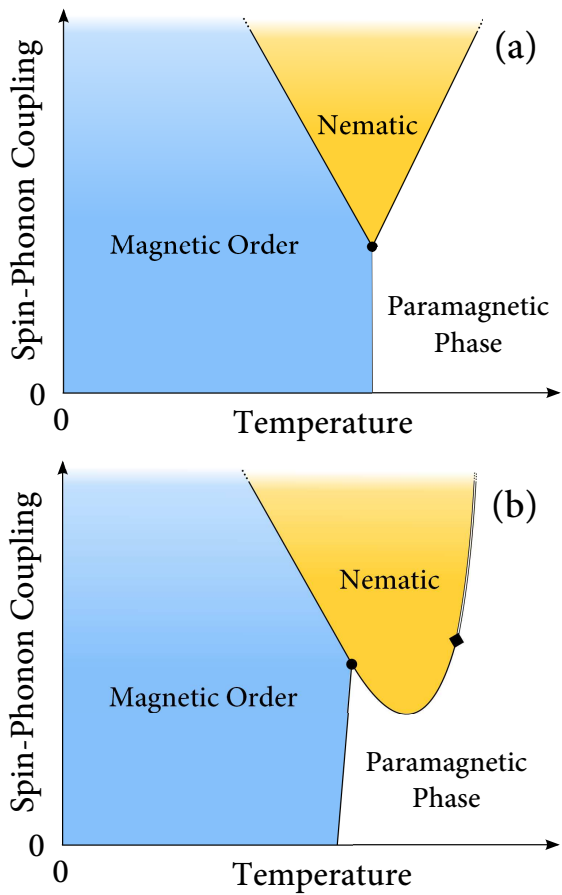}
   \caption{
   Schematic phase diagram in the plane of temperature versus spin-lattice coupling based on (a) mean-field theory plus Gaussian fluctuations and (b) XY critical scaling (see Sec.~\ref{sec:elastic-mean-field}). The dot in the phase diagrams is the multicritical point connecting the paramagentic, nematic, and magnetically ordered phases. The solid diamond in (b) is a tricritical point; the phase transition become first order for stronger coupling, while it remains mean-field-like for weaker coupling.
   }
   \label{fig:intro:phdiag}
\end{figure}

The low temperature phase in both models has four possible domains, describable by a $Z_4$ order parameter, breaking simultaneously both time-reversal symmetry and lattice rotation symmetry from tetragonal to orthorhombic.  Monte Carlo simulations show that the transition to the $Z_4$ state occurs in a single, continuous, transition.  We argue this is also a consistent possibility according to renormalization group theory, and predict that the continuous transition is in the 3d XY universality class.  However, if we consider the phonon degrees of freedom, another possibility emerges, since the lattice energy is sensitive to the orthorhombic distortion: spin-lattice coupling may induce an intermediate temperature phase breaking lattice rotations but not time-reversal.  This would be a ``nematic'' state in modern parlance, or simply a paramagnetic orthorhombic state in more conventional terms.  Using the theory of critical phenomena, we argue that spin-lattice coupling may indeed induce a nematic phase, but that this occurs only beyond some non-zero threshold coupling strength.  Hence, the appearance of a nematic phase implies the presence of strong spin-lattice coupling.  A system in which spin-lattice coupling is close to the threshold value for the onset of nematic order is governed by a universal {\em multicritical} point, as shown in Figure~\ref{fig:intro:phdiag}.

The remainder of the paper is organized as follows.  In Sec.~\ref{sec:model}, we present the uniform and staggered tetragonal models.  Sec.~\ref{sec:resA} describes the analysis of the simpler uniform tetragonal model: its classical ground states, the effects of thermal and quantum fluctuations, and its phase transitions.  Next, in Sec.~\ref{sec:resB}, we consider the staggered sublattice model, and show that it exhibits features relevant to Ba$_2$NaOsO$_6$ already at the classical level, but does not show a nematic phase.  Spin-lattice coupling is described in Sec.~\ref{ssec:resB:gl}.  Finally, we review the main results and discuss implications for experiments in Sec.~\ref{sec:discussions-summary}.  Several appendices give additional details of calculations to support the main text.

\section{Model}\label{sec:model}

In this section, we introduce the models we study in this paper. In Sec.~\ref{ssec:model:cubic} we first introduce a general model with NN interaction that is allowed by cubic symmetry. General models with symmetry-allowed NN interactions considering tetragonal symmetry are introduced in Sec.~\ref{ssec:model:tetra} and Sec.~\ref{sec:stagg-tetr-case}.

\subsection{Cubic}\label{ssec:model:cubic}

When the fcc lattice has cubic symmetry, assuming pairwise interactions amongst $S=1/2$ spins, only two different interactions are allowed. The Hamiltonian is given by
\begin{eqnarray}
H_\text{cb} &=& \frac{J_1}2 \sum_{{\bf R},{\bm \delta}\in \text{NN}} {\bf S}_{\bf R}\cdot{\bf S}_{{\bf R}+{\bm \delta}} \nonumber\\
       &&\qquad + \frac{J_3}2 \sum_{{\bf R},{\bm \delta}\in \text{NN}} ({\bf S}_{\bf R}\cdot{\boldsymbol \delta})({\bf S}_{{\bf R}+{\boldsymbol \delta}}\cdot{\boldsymbol \delta}).\nonumber\\ \label{eq:Hcubic}
\end{eqnarray}
Here, ${\bf S}_{\bf R}=(S_{\bf R}^{(x)},S_{\bf R}^{(y)},S_{\bf R}^{(z)})$ is the spin operator for $S=1/2$ spins at site $\bf R$ and $\bm \delta$ is the vector connecting nearest-neighbor (NN) sites. The sum for $\bf R$ is taken over all the sites and that for $\bm \delta$ is over all NN sites
\begin{eqnarray}
{\boldsymbol \delta}&=&(\pm\frac1{\sqrt2},\pm\frac1{\sqrt2},0),\;(\pm\frac1{\sqrt2},0,\pm\frac1{\sqrt2}),\\ \nonumber
&&\;(0,\pm\frac1{\sqrt2},\pm\frac1{\sqrt2}).
\end{eqnarray}

\subsection{Uniform tetragonal case}\label{ssec:model:tetra}

When the symmetry is reduced to tetragonal -- e.g., the lattice is shortened/elongated along $z$ axis -- the general spin model with NN interactions consist of seven interactions:
\begin{eqnarray}
H_\text{Tetra}= H^{XY} + H^Z \label{eq:model:Htetra}
\end{eqnarray}
with
\begin{eqnarray}
{\cal H}^{XY} &=& \frac12 \sum^{XY}_{{\bf R}, {\boldsymbol \delta}} J_1 ( S^x_{{\bf R}}S^x_{{\bf R}+{\boldsymbol\delta}} + S^y_{{\bf R}}S^y_{{\bf R}+{\boldsymbol\delta}} )\\ \nonumber
&&+ J_2 S^z_{{\bf R}}S^z_{{\bf R}+{\boldsymbol\delta}}
+ J_3 ({\bf S}_{\bf R}\cdot{\boldsymbol \delta})({\bf S}_{{\bf R}+{\boldsymbol \delta}}\cdot{\boldsymbol \delta}) \label{eq:model:Htetra_xy}
\end{eqnarray}
and
\begin{eqnarray}
{\cal H}^{Z}&=& \frac12 \sum^{Z}_{{\bf R}, {\boldsymbol \delta}} K_1 ( S^x_{{\bf R}}S^x_{{\bf R}+{\boldsymbol\delta}} + S^y_{{\bf R}}S^y_{{\bf R}+{\boldsymbol\delta}} ) + K_2 S^z_{{\bf R}}S^z_{{\bf R}+{\boldsymbol\delta}}\\ \nonumber
&&+ K_3 ({\bf S}_{\bf R}\cdot{\boldsymbol \delta})({\bf S}_{{\bf R}+{\boldsymbol \delta}}\cdot{\boldsymbol \delta}) + K_4({\bf S}_{\bf R}\cdot\tilde{\boldsymbol \delta})({\bf S}_{{\bf R}+{\boldsymbol \delta}}\cdot\tilde{\boldsymbol \delta}). \label{eq:model:Htetra_z}
\end{eqnarray}
Here, the sums for ${\bm \delta}$ in $H^{XY}$ ($H^{Z}$) are taken only for the bonds in (out of) the $XY$ planes, i.e., ${\bm \delta}\cdot{\bf z}=0$ (${\bm \delta}\cdot{\bf z}\ne 0$) with $\bf z$ being the unit vector along $z$ axis.
In $H^{Z}$, $\hat{\boldsymbol \delta}$ is the projection of $\bm \delta$ on to the $XY$ plane,
\begin{eqnarray}
\hat{\boldsymbol \delta} = (\delta_x,\delta_y,0),
\end{eqnarray}
with $\delta_\alpha$ ($\alpha=x,y,z$) being the $\alpha$ component of $\bm \delta$.
The magnetic behavior of this model is studied in Sec.~\ref{sec:resA}.

\subsection{Staggered tetragonal case}
\label{sec:stagg-tetr-case}

Tetragonal anisotropy may also occur via the presence of inequivalent sublattices.
We consider the case shown in Fig.~\ref{fig:intro:fcc}(b), where the local environment for ions alternates along the $z$ axis; we call the two sublattices $A$ and $B$.
Inspired by a previous study,~\cite{Chen2010} in this two-sublattice model, we suppose there is no $C_4$ rotation symmetry (i.e. no four-fold rotation symmetry in a single xy plane) but instead only a $C_4$ screw axis parallel to $z$ axis.

The Hamiltonian for the two-sublattice model is given by
\begin{eqnarray}
H = H_{A} + H_{B} + H_{AB}, \label{eq:model:H2sub}
\end{eqnarray}
where
\begin{widetext}
\begin{eqnarray}
H_A    &=& \sum_{\substack{\delta_z=0\\{\bf R}\in A}} \frac{J_1}2 S^x_{\bf R} S^x_{{\bf R}+{\bm \delta}} + \frac{J_1^\prime}2 S^y_{\bf R} S^y_{{\bf R}+{\bm \delta}} + \frac{J_2}2 S^z_{\bf R} S^z_{{\bf R}+{\bm \delta}}+\frac{J_3}2\delta_x\delta_y(S^x_{\bf R} S^y_{{\bf R}+{\bm \delta}}+S^y_{\bf R} S^x_{{\bf R}+{\bm \delta}})\\
H_B    &=& \sum_{\substack{\delta_z=0\\{\bf R}\in B}} \frac{J_1^\prime}2 S^x_{\bf R} S^x_{{\bf R}+{\bm \delta}} + \frac{J_1}2 S^y_{\bf R} S^y_{{\bf R}+{\bm \delta}} + \frac{J_2}2 S^z_{\bf R} S^z_{{\bf R}+{\bm \delta}}+\frac{J_3}2\delta_x\delta_y(S^x_{\bf R} S^y_{{\bf R}+{\bm \delta}}+S^y_{\bf R} S^x_{{\bf R}+{\bm \delta}})\\
H_{AB} &=& \sum_{\substack{\delta_z\ne0\\{\bf R}\in B}} (K_1\delta_x^2+K_1^\prime\delta_y^2) S^x_{\bf R} S^x_{{\bf R}+{\bm \delta}} + (K_1^\prime\delta_x^2+K_1\delta_y^2) S^y_{\bf R} S^y_{{\bf R}+{\bm \delta}} + \frac{K_2}2 S^z_{\bf R} S^z_{{\bf R}+{\bm \delta}}\nonumber\\
&&\qquad\quad +K_3^\prime\left\{\delta_z\delta_x(S^x_{\bf R} S^z_{{\bf R}+{\bm \delta}}+S^z_{\bf R} S^x_{{\bf R}+{\bm \delta}}) + (x\to y)\right\} + K_4^\prime\left\{\delta_z\delta_x(S^x_{\bf R} S^z_{{\bf R}+{\bm \delta}}-S^z_{\bf R} S^x_{{\bf R}+{\bm \delta}}) + (x\to y)\right\}.\\
H_{AB} &=& \sum_{\substack{\delta_z\ne0\\{\bf R}\in B}} (K_1\delta_x^2+K_2\delta_y^2) S^x_{\bf R} S^x_{{\bf R}+{\bm \delta}} + (K_2\delta_x^2+K_1\delta_y^2) S^y_{\bf R} S^y_{{\bf R}+{\bm \delta}} + \frac{K_3}2 S^z_{\bf R} S^z_{{\bf R}+{\bm \delta}}\nonumber\\
&&\qquad\quad +K_4\left\{\delta_z\delta_x(S^x_{\bf R} S^z_{{\bf R}+{\bm \delta}}+S^z_{\bf R} S^x_{{\bf R}+{\bm \delta}}) + (x\to y)\right\} + K_5\left\{\delta_z\delta_x(S^x_{\bf R} S^z_{{\bf R}+{\bm \delta}}-S^z_{\bf R} S^x_{{\bf R}+{\bm \delta}}) + (x\to y)\right\}.
\end{eqnarray}
\end{widetext}
Here, $H_A$ ($H_B$) contains the bonds within one $xy$ plane, and $H_{AB}$ gives the interactions between the spins on different layers.  The notation
$(x\to y)$ denote a term with same form as the preceding one but with the $x$ component replaced by the $y$ component. 

A notable difference of the Hamiltonian in Eq.~(\ref{eq:model:H2sub}) from the Hamiltonian in Eq.~(\ref{eq:model:Htetra}) is the presence of Ising-type interactions that alternate from layer to layer, e.g., $J_1$ and $J_1^\prime$.
Focusing on the role of these interactions, we mainly study a simplified model
\begin{eqnarray}
H_{2S} &=& \frac{J}2\sum_{{\bf R},{\bm \delta}}{\bf S}_{\bf R}\cdot{\bf S}_{{\bf R}+{\bm \delta}}\nonumber\\
&+& \frac{J^\prime}2\{\sum_{\substack{{\bf R}\in A,\\\delta_z=0}} S^x_{\bf R}\cdot S^x_{{\bf R}+{\bm \delta}}
+ \sum_{\substack{{\bf R}\in B,\\\delta_z=0}} S^y_{\bf R}\cdot S^y_{{\bf R}+{\bm \delta}}\}. \label{eq:model:H2S}
\end{eqnarray}
The results are presented in Sec.~\ref{sec:resB}.

\section{Tetragonal Equivalent-Sublattice Model} \label{sec:resA}

In this section, we study the model in Eq.~(\ref{eq:model:Htetra}).
First, we investigate the classical ground state of the model by the Luttinger-Tisza method.~\cite{Luttinger1946} We then turn to the effects of fluctuations, focusing on a particular ordered phase for concreteness. Thermal and quantum fluctuations are studied in Secs.~\ref{ssec:resA:tfluc} and Sec.~\ref{ssec:resA:qfluc}, respectively.

\subsection{Ground state} \label{ssec:resA:gs}

\begin{figure}
\begin{center}
\includegraphics[width=.85\linewidth]{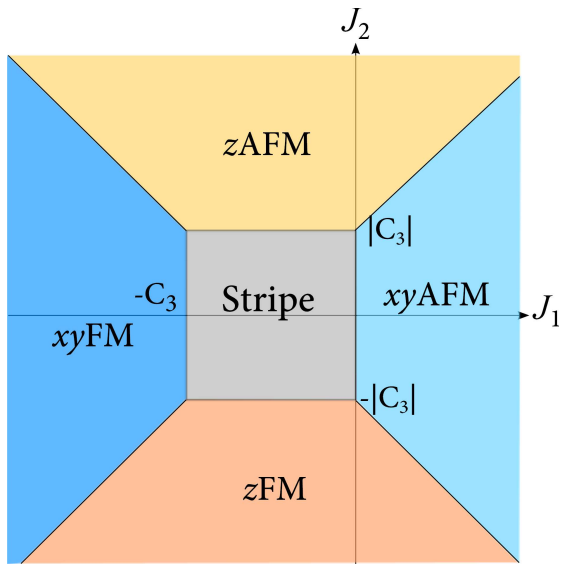}
\end{center}
\caption{
Ground state phase diagram of the Hamiltonian in Eq.~(\ref{eq:model:Htetra}) in the 2D limit obtained by the Luttinger-Tisza method. The phase boundaries are given in Table~\ref{tab:resA:gs}. See the text for details.
}\label{fig:resA:2ddiag}
\end{figure}

\begin{table*}
   \begin{tabular}{c|c}
   \hline
   \hspace{30pt}Order\hspace{30pt}  & \hspace{300pt} \\
   \hline
   $xy$FM   & $c_1 < \min(-c_3,0)$,   $c_2 < -c_1-\frac{c_3}2$, $c_2 > c_1+\frac{c_3}2$ \\
   $xy$AFM & $c_1 > \max(0,-c_3)$,   $c_2 > -c_1-\frac{c_3}2$, $c_2 < c_1+\frac{c_3}2$ \\
   $z$FM    & $c_2 < -\frac{|c_3|}2$, $c_2 < -c_1-\frac{c_3}2$, $c_2 < c_1+\frac{c_3}2$ \\
   $z$AFM & $c_2 > \frac{|c_3|}2$,  $c_2 > -c_1-\frac{c_3}2$, $c_2 > c_1+\frac{c_3}2$ \\
   Stripe    & $0 > c_3(c_1+c_3)$, $c_2 < \frac{|c_3|}2$ \\
   \hline
   \end{tabular}
   \caption{
   Regions in the phase space for each phases shown in Fig.~\ref{fig:resA:2ddiag} to be the ground state.
   The conditions are obtained by Luttinger-Tisza method.
   }
   \label{tab:resA:gs}
\end{table*}

\subsubsection{Independent layers}
\label{sec:independent-layers}

We first consider the magnetic phase diagram in absence of interlayer coupling, $K_i$ in Eq.~(\ref{eq:model:Htetra}). The phase diagram for the independent layers is shown in Fig.~(\ref{fig:resA:2ddiag}) and the phase boundaries in Table~\ref{tab:resA:gs}. When $J_1$ is dominant and ferromagnetic, the ground state is ferromagnetic with spins pointing in $xy$ plane. The phase is denoted as $xy$FM in the phase diagram. On the other hand, when $J_1$ is dominant but antiferromagnetic, a Neel state with spins pointing in $xy$ plane become the ground state ($xy$AFM in Fig.~\ref{fig:resA:2ddiag}).   An interesting feature of these phases is an accidental U(1) degeneracy: despite the absence of U(1) symmetry in the Hamiltonian, the classical ground state energy is independent of the direction of the moments in the $xy$ plane.
The U(1) degeneracy is a feature of dominant $J_1$ interactions.  If instead $J_2$ is dominant, spins align along the $z$ axis, either ferromagnetically ($J_2<0$) or antiferromagnetically ($J_2>0$) depending upon the sign of interaction.

When $J_3=0$, these four phases meet at $J_1=J_2=0$. On the other hand, with infinitesimally small $J_3\ne0$, a stripe order appears in the competition region where the four phases meet (see Fig.~\ref{fig:resA:2ddiag}). In this stripe phase, the moments align ferromagnetically along one of the bonds, and antiferromagnetically along the orthogonal bonds.  They lie in the $xy$ plane and point along the $\langle110\rangle$ direction.  In the stripe phase the axis of the moments and the wave-vector are related. When $J_3>0$, the direction of moments and wave-vector are parallel, while when $J_3<0$, the direction of moments and the wavevector are orthogonal.  In either of these cases, the ground state is four-fold degenerate.

\subsubsection{Three-dimensional coupling}
\label{sec:three-dimens-coupl}

We next consider how the interlayer couplings modify the phase diagram in Fig.~\ref{fig:resA:2ddiag}. We here focus on the $xy$ FM state and study how it changes with $K_1$ and $K_2$. The phase diagrams for the other regions are described in Appendix~\ref{sec:appA}.

\begin{figure}
\begin{center}
\includegraphics[width=.85\linewidth]{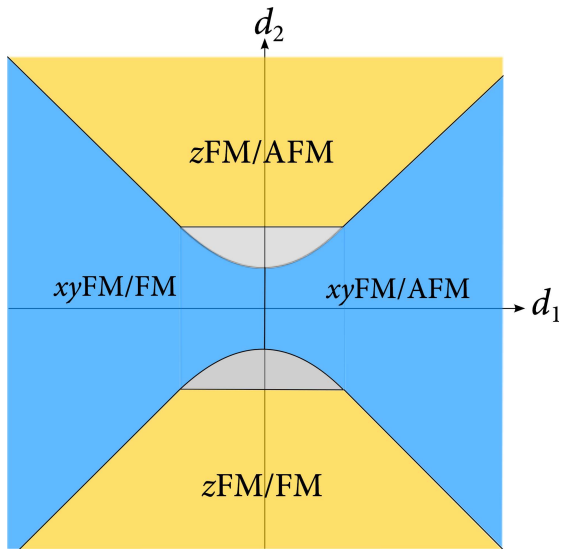}
\end{center}
\caption{
Ground state phase diagram of the Hamiltonian in Eq.~(\ref{eq:model:Htetra}) in the region where in-plane ferromagnetic ($xy$FM) orders become the ground state in the 2d limit. The phase diagram is obtained by the Luttinger-Tisza method. The gray region denotes regions where we could not determine the ground state. See the text for details. 
}\label{fig:resA:3ddiag}
\end{figure}

Figure~\ref{fig:resA:3ddiag} shows the ground state phase diagram of Hamiltonian in Eq.~(\ref{eq:model:Htetra}) with ($K_1$,$K_2$) in the regime so that the $xy$FM state appears in the independent layer model of Fig.~\ref{fig:resA:2ddiag}. In this region, introduction of $K_1<0$ selects ferromagnetic ordering of the 2d planes ($xy$FM/FM state), while antiferromagnetic stacking of the ferromagnetic layers is selected for $K_1>0$ ($xy$FM/AFM state). The phase boundary of the two phases are located at $K_1=0$.

Introducing $K_2$ gives rise to phase competition between Ising and $xy$ orders.   For sufficiently large negative $K_2<0$, spin align ferromagnetically and parallel to the $z$ axis ($z$FM/FM state). As shown in the lower side of Fig.~\ref{fig:resA:3ddiag}, this occurs when
\begin{eqnarray}
K_2 < -|K_1|-\frac12(J_2-J_1-\frac12J_3), \label{eq:resA:xyFMb1}
\end{eqnarray}
while the $xy$FM orders become the ground state for larger but negative $K_2$.

For $J_2<0$, the situation is somewhat more complicated.  In particular, when 
\begin{eqnarray}
 2J_2|K_1|-J_2(J_1+\frac{J_3}2) < |K_2|^2 < J_2^2, \label{eq:resA:xyFMb2}
\end{eqnarray}
the Luttinger-Tisza method cannot verify either an antiferromagnetic or ferromagnetic ground state.  A definitive identification of the ground state in this region requires further analysis.  Consequently, this region is left blank in Figure~\ref{fig:resA:3ddiag}.  In principle, the Luttinger-Tisza method gives the minimum region of stability for a given magnetic order.  Hence, it is expected that at least some portion of this area is, in reality, covered by the magnetic orders surrounding the region. For large positive $K_2>0$, the ground state is given by antiferromagnetic stacking of the $z$FM layers ($z$FM/AFM), and otherwise the phase diagram for $K_2>0$ mirrors that for $K_2<0$, as shown in Fig.~\ref{fig:resA:3ddiag}.

\subsection{Thermal fluctuation} \label{ssec:resA:tfluc}

In Sec.~\ref{ssec:resA:gs}, we found a wide region of magnetic phases with spins oriented within the $xy$ plane, in several of which, despite the magnetic anisotropy of the Hamiltonian, a U(1) degeneracy under rotations within the plane is recovered. However, as there is no symmetry that protects this degeneracy, it is expected to be lifted by perturbations. Indeed, as we will show in the rest of this section, both thermal and quantum fluctuations lift this degeneracy selecting $\langle110\rangle$ directions.

In this section, taking the $xy$FM/AFM phases as a prototypical example, we consider how thermal fluctuations modify this classical degeneracy. In the classical limit, the stacking along $z$ axis does not matter, since the transformation
\begin{eqnarray}
{\bf S}_{\bf R} \to \left\{
\begin{array}{ll}
 {\bf S}_{\bf R} & (\text{if} \quad R_z = \frac{2n}{\sqrt2})   \\
-{\bf S}_{\bf R} & (\text{if} \quad R_z = \frac{2n+1}{\sqrt2})
\end{array}
\right.
\end{eqnarray}
changes the sign of $K_1$ in the model in Eq.~(\ref{eq:model:Htetra}). Here, $n$ is an integer.

In Sec.~\ref{ssec:resA:1loop}, we first consider the one-loop correction to the classical spin model due to thermal fluctuations at low temperature.  We show that magnetic anisotropy is induced by $J_3$.  We then verify and study anisotropy in higher temperature by a classical MC simulation in Sec.~\ref{ssec:resA:mc}.

\subsubsection{Classical spin-wave theory}\label{ssec:resA:1loop}

Let us first consider the effect of thermal fluctuations in the region $T\ll T_c$. In this limit, each spin is close to its ground state orientation. Hence, we may expand in small fluctuations $\delta{\bf S}_{\bf R}$  around  their average magnetic moment ${\bf M}_{\bf R}$,
\begin{eqnarray}
{\bf S}_{\bf R} = {\bf M}_{\bf R} + \delta{\bf S}_{\bf R}.
\end{eqnarray}
The distribution function at $T\ll T_c$ can be approximated as~\cite{Henley1987}
\begin{eqnarray}
Z          &\sim& \int d{\bf M} Z({\bf M})\\
Z({\bf M}) &\propto&    \int \prod_{{\bf k}\ne{\bf 0}}d(\delta{\bf S}_{\bf k}) \exp(-\beta \sum_{{\bf k}\ne{\bf 0}}\delta{\bf S}_{\bf k}\cal{J}({\bf k})\delta{\bf S}_{\bf k}).\nonumber\\ \label{eq:1loop_Z}
\end{eqnarray}
where $Z({\bf M})$ is the partial distribution function with the net magnetic moment pointing along $\bf M$, and $\delta{\bf S}_{\bf k}$ is the Fourier transform of $\delta{\bf S}_{\bf R}$. Similarly, $\cal{J}({\bf k})$ is the Fourier transform of the $3\times3$ interaction matrix $J({\bf R},{\bf R}^\prime)$, where general form of Hamiltonian is given by
\begin{eqnarray}
H = \frac12\sum_{{\bf R},{\bf R}^\prime}{}^t{\bf S}_{\bf R}\,J({\bf R},{\bf R}^\prime)\,{\bf S}_{{\bf R}^\prime}.
\end{eqnarray}
As we focus on the low $T$ limit, we may approximate
\begin{eqnarray}
\delta{\bf S}_{\bf R}\cdot {\bf M} = 0.
\end{eqnarray}

Calculating the Gaussian integral in Eq.~(\ref{eq:1loop_Z}) with the orthogonal constraint for $\delta{\bf S}_{\bf R}$ gives free energy for the ordered state with net magnetic moment $\bf M$ as
\begin{eqnarray}
F({\bf M}) &=&       -T\log Z({\bf M})\\
     &\propto& \frac12 \sum_{\bf k} \log \left[{\cal J}_1({\bf k})-{\cal J}_1^\prime({\bf k})\cos2\theta\right] + \log {\cal J}_2({\bf k})\nonumber\\.
\end{eqnarray}
\begin{widetext}
Here,
\begin{eqnarray}
{\cal J}_1({\bf k})        &=& 2(2J_1+J_3)\cos\frac{k_x}{\sqrt2}\cos\frac{k_y}{\sqrt2}  + 4K_1 \cos\frac{k_z}{\sqrt2}[\cos k_x + \cos k_y] - 2(2J_1+J_3+4K_1)\\
{\cal J}_1^\prime({\bf k}) &=& -4J_3\sin\frac{k_x}{\sqrt2}\sin\frac{k_y}{\sqrt2}\\
{\cal J}_2({\bf k})        &=& 4J_2\cos\frac{k_x}{\sqrt2}\cos\frac{k_y}{\sqrt2} + 4K_2 \cos\frac{k_z}{\sqrt2}[\cos k_x + \cos k_y] - 4(J_2+2K_2)
\end{eqnarray}
\end{widetext}
and
\begin{eqnarray}
{\bf M}=(\cos\theta,\sin\theta,0).
\end{eqnarray}
The result clearly shows that $F({\bf M})$ depends upon the angle $\theta$, indicating that thermal fluctuations break the U(1) degeneracy of the ground state.

The location of the minimum is determined by the vanishing derivative with respect to $\theta$, which is given by
\begin{eqnarray}
\partial_\theta F({\bf M}) = -\frac{T\sin4\theta}4\sum_{\bf k}\frac{{\cal J}_1^\prime{}^2({\bf k})}{{\cal J}_1{}^2({\bf k})-{\cal J}_1^\prime{}^2({\bf k})\sin^22\theta}.
\end{eqnarray}
As $|{\cal J}_1({\bf k})| > |{\cal J}_1^\prime({\bf k})|$ for ${\bf k}\ne {\bf 0}$, when $c_3\ne0$, the sum over $\bf k$ is always positive and finite. Hence, $\sin4\theta=0$ gives the location of the maxima/minima; $\theta=\frac{(2n+1)\pi}4$ gives the minimum of $F({\bf M})$. Thus, thermal fluctuations favor $\langle110\rangle$ magnetic anisotropy irrespective of the sign of $J_3$, reducing the U(1) symmetry of the ordered phase to $Z_4$.   We observe that if $J_3=0$, $F({\bf M})$ becomes independent of $\bf M$. This is natural as $J_1$ and $J_2$ do not break the U(1) symmetry.

\subsubsection{Monte Carlo simulation}\label{ssec:resA:mc}

\begin{figure}
\includegraphics[width=.96\linewidth]{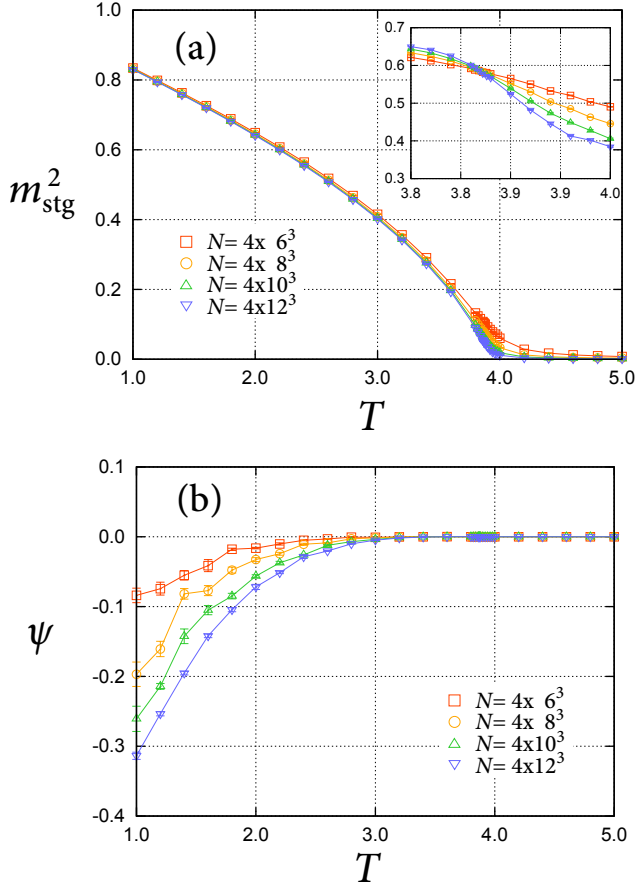}
\caption{
	Classical Monte Carlo simulation of the Hamiltonian in Eq.~(\ref{eq:model:Htetra}). Temperature dependence of (a) $m^{(xy)}_\text{stg}{}^2$ and (b) $\psi$. The inset in (a) shows the Binder parameter for $m^{(xy)}_\text{stg}$.
}\label{fig:mc_tetra}
\end{figure}

This spin anisotropy, induced by thermal fluctuations, appears over a wide range of temperatures. This is confirmed by numerical simulation using classical MC method. Fig.~\ref{fig:mc_tetra} shows a result of the MC simulation for $J_1=-1$, $J_3=-1$, and $K_1=1$; the other parameters are set to zero.

Figure~\ref{fig:mc_tetra}(a) shows temperature dependence of staggered magnetization $m^{(xy)}_\text{stg}{}^2=(m_\text{stg}^x)^2+(m_\text{stg}^y)^2$ calculated for different system sizes. Here,
\begin{eqnarray}
(m_\text{stg}^\alpha)^2 = \left<\frac1{N^2}\left(\sum_{\bf R}(-1)^{\sqrt2 R_z}S_{\bf R}^{(\alpha)}\right)^2\right>,\label{eq:mstg_xy}
\end{eqnarray}
with $\alpha=x,y,z$ is the $\alpha$th component of the staggered magnetization. The result shows an increase of $m^2$ below $T\sim3.9$, indicating magnetic phase transitions. The critical temperature $T_c$ is estimated from the Binder parameter~\cite{Binder1981} for $m^{(xy)}_\text{stg}$. The inset in Fig.~\ref{fig:mc_tetra}(a) shows the Binder parameter calculated in the vicinity of $T_c$. The result shows a monotonic decrease with increasing temperature with a crossing at $T=T_c=3.868(6)$. This is an indicative of second order phase transition.

To study the magnetic anisotropy in the ordered phase at $T<T_c$, we calculated
\begin{eqnarray}
\psi &=& \left< \frac1N\sum_{\bf R} (S_{\bf R}^{(x)}{}^2 +S_{\bf R}^{(y)}{}^2)^2\cos4\theta_{\bf R}\right>\\
             &=& \left<\frac1N\sum_{\bf R}(S_{\bf R}^{(x)}{}^2-S_{\bf R}^{(y)}{}^2)^2-4S_{\bf R}^{(x)}{}^2S_{\bf R}^{(y)}{}^2\right>.
\end{eqnarray}
The parameter $\psi$ becomes positive when the magnetic moments preferentially point along the $\langle100\rangle$ directions while it becomes negative when the spins point along $\langle110\rangle$.

On lowering temperature below $T_c$, $\psi$ gradually deviates from zero to negative value $\psi<0$, as shown in Fig.~\ref{fig:mc_tetra}(b). This indicates that spins align along the $\langle 110\rangle$ directions, consistent with the one-loop analysis above. However, in contrast to $m^2$, $\psi$ does not show a sharp increase at $T_c$. Also, $\psi$ shows a strong finite size effect even at temperatures $T\ll T_c$.  To understand this, we note that in the XY model, the cubic-anisotropy term is a dangerously irrelevant parameter,~\cite{Jose1977,Blankschtein1984,Caselle1998,Oshikawa2000,Carmona2000} i.e. irrelevant at the critical point but relevant for $T<T_c$.  This implies the existence of a second length scale $\Lambda$ which is {\it larger} than the usual correlation length. Hence, when the system size is small, the large length $\Lambda$ may obscure the anisotropic behavior. Indeed, similar behavior was reported in a recent MC simulation of a 3D XY model with $Z_q$ anisotropy.~\cite{Lou2007}

\begin{figure}
\includegraphics[width=.9\linewidth]{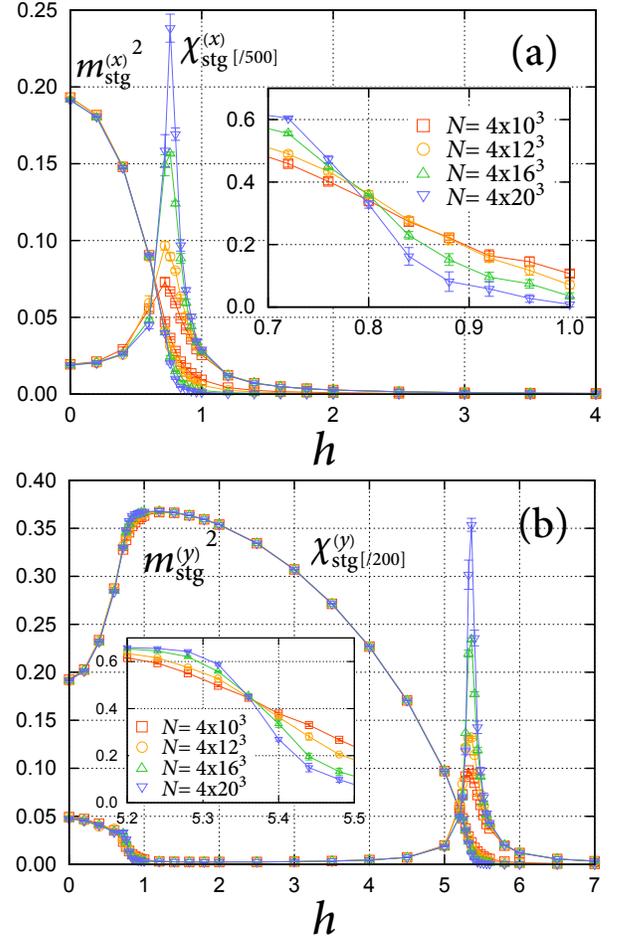}
\caption{
Magnetic field dependence of the staggered magnetization and susceptibility along (a) $x$ and (b) $y$ axes at $J_1=0.5$, $J_3=-1$, and $K_1=0.6$.
The insets in (a) and (b) shows the field dependence of the Binder parameter for staggered magnetization along the $x$ and $y$ axes, respectively. 
}\label{fig:mc_mag}
\end{figure}

Next, we study the effect of external magnetic field, which leads, we will see, to features that reflect the fluctuation-induced anisotropy. Fig.~\ref{fig:mc_mag} shows the field dependence of the staggered magnetization along different crystal axes at $T=1$. Here, we used parameters $J_1=0.5$, $J_3=-1$, and $J_1=0.6$, and the external field was applied along the $[100]$ direction. With increasing field, the staggered magnetization along the $x$ axis vanishes first, which is seen by the rapid decrease of the magnetization to zero at $H_{c1}=0.78(4)$ [Fig.~\ref{fig:mc_mag}(a)].  This phase transition is also observed in the magnetic susceptibility $\chi^{(x)}$. Here, we calculated $\chi^{(\alpha)}$ ($\alpha=x,y,z$), by the fluctuation formula,
\begin{eqnarray}
\chi^{(\alpha)}_\text{stg} = \frac{N}{T}\{(m^{(\alpha)}_\text{stg})^2 -({\bar m}^{(\alpha)}_\text{stg})^2\}
\end{eqnarray}
with
\begin{eqnarray}
{\bar m}^{(\alpha)}_\text{stg} = \left<\frac1{N}\left|\sum_{\bf R}(-1)^{\sqrt2 R_z}S_{\bf R}^{(\alpha)}\right|\right>.
\end{eqnarray}
The result is also shown in Fig.~\ref{fig:mc_mag}(a), which shows diverging behavior at $H_{c1}$.  With further increase of the field, the $y$ component of staggered magnetization vanishes via a second transition at $H_{c2}=5.36(2)$; a similar divergence of susceptibility is also observed [Fig.~\ref{fig:mc_mag}(b)]. The critical fields are estimated from the Binder parameters for $m_\text{stg}^{(x)}$ and $m_\text{stg}^{(y)}$, which are shown in the insets.

\begin{figure}
\includegraphics[width=.9\linewidth]{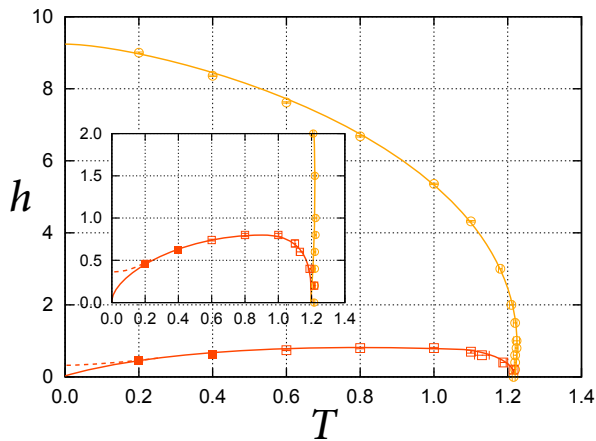}
\caption{
Temperature dependence of $H_{c1}$ and $H_{c2}$ for Hamiltonian in Eq.~(\ref{eq:model:Htetra}) at $J_1=0.5$, $J_3=-1$, and $K_1=0.6$. The inset shows an enlarged view of the low field region. The solid lines are guides for the eye, and the dotted line is a sketch of the expected phase boundary in the presence of quantum fluctuations.
}\label{fig:mc_diag}
\end{figure}

Figure~\ref{fig:mc_diag} shows the $H-T$ phase diagram for $J_1=0.5$, $J_3=-1$, and $K_1=0.6$. The upper boundary $H_{c2}(T)$ monotonically decreases with increasing temperature in a conventional fashion. On the other hand,  $H_{c1}(T)$ shows a non-monotonic behavior with an increase with increasing temperature for $T\ll T_c$ and a maximum around $T=0.8$.  This behavior is characteristic of thermal-fluctuation induced anisotropy, since the spin anisotropy weakens at low temperature, hence $H_{c1}(T)$ approaches $H_{c1}\to 0$ as $T\to0$. The non-monotonic behavior can therefore be used as an experimental diagnostic of thermal fluctuation-induced anisotropy.

Lastly, we briefly mention the behavior of the $xy$FM/FM state in mangetic field. The above argument on magnetic anisotropy remains applicable since the $xy$FM/AFM case can be related using a transformation mentioned in the begining of this section.   This correspondance is, however, no longer valid in the presence of the external magnetic field. In this case, it is expected that there is a single transition at $H_c$, above which the net magnetic moment aligns parallel to the external field. As the anisotropy is driven by the fluctuations, the temperature dependence of $H_c$ is expected to be non-monotonic, similar to $H_{c1}$ shown in Fig.~\ref{fig:mc_diag}.

\subsection{Quantum fluctuations}\label{ssec:resA:qfluc}

We next focus on the effect of quantum fluctuations, which are expected to be substantial and dominate at low temperature for $S=1/2$ quantum spins.  By a spin-wave analysis, we show that quantum fluctuations act similarly to thermal fluctuations, also giving rise to four-fold magnetic anisotropy favoring the $\langle110\rangle$ directions.

In the antiferromagnetic spin wave theory, the ground state energy is given by
\begin{eqnarray}
E_\text{GS} = E_{cl} + N(2J_1+J_3-4K_1) +\frac12 \sum_{\alpha,{\bf k}} \omega_{\alpha,{\bf k}},
\end{eqnarray}
where $E_{cl}$ is the classical ground state energy and $\omega_{\alpha,{\bf k}}$ is the spin-wave dispersion for $\alpha$ mode ($\alpha=\pm$). The latter two terms give the quantum correction to the energy. In the current model, $\omega_{\pm,{\bf k}}$ is given by
\begin{widetext}
\begin{eqnarray}
\omega_{\pm,\bf k}&=& \sqrt{\omega_0\{\omega_0+a_\pm({\bf k})+b({\bf k})\sin2\theta\}}\\
\omega_0   &=& -2J_1-J_3+4K_1,
\end{eqnarray}
where
\begin{eqnarray}
a_\pm({\bf k}) &=& (2J_1+J_3)\cos\frac{k_x}{\sqrt2}\cos\frac{k_y}{\sqrt2} \pm 2K_1\cos\frac{k_z}{\sqrt2}(\cos\frac{k_x}{\sqrt2}+\cos\frac{k_y}{\sqrt2})\\
b({\bf k}) &=& J_3\sin\frac{k_x}{\sqrt2}\sin\frac{k_y}{\sqrt2}.
\end{eqnarray}
The minimum of $E_\text{GS}$ with respect to $\theta$ is determined from the derivative $\partial_\theta E_\text{GS}$, which is given by
\begin{eqnarray}
\partial_\theta E_\text{GS} &=& \frac12\sum_{\alpha,{\bf k}} \frac{\omega_0 b({\bf k})\cos2\theta}{\sqrt{\omega_0\{\omega_0+a_\alpha({\bf k})+b({\bf k})\sin2\theta\}}} \\
    &=& -\frac{\omega_0^{1/2}}4\sum_{\alpha,{\bf k}} \frac{b({\bf k})\cos2\theta}{\sqrt{(\omega_0+a_\alpha({\bf k}))^2-b^2({\bf k})\sin^22\theta}} \left\{\sqrt{\omega_0+a_\alpha({\bf k})+b({\bf k})\sin2\theta}-\sqrt{\omega_0+a_\alpha({\bf k})-b({\bf k})\sin2\theta}\right\}.\label{eq:qcorr}
\end{eqnarray}
\end{widetext}
Expanding the last bracket in Eq.~(\ref{eq:qcorr}) in $b({\bf k})\sin2\theta$ gives
\begin{eqnarray}
\partial_\theta E_\text{GS} = -\frac18\sum_{\alpha,{\bf k}} E_{\alpha,{\bf k}}\,b^2({\bf k})\sin4\theta,\label{eq:dE_dth}
\end{eqnarray}
with
\begin{eqnarray}
E_{\alpha,{\bf k}} = E^{(0)}_{\alpha,{\bf k}} \sum_{n=\text{odd}} f_n(\frac{b({\bf k})}{\sqrt{\omega_0+a_\alpha({\bf k})}}).
\end{eqnarray}
Here, the sum is for all the odd natural numbers,
\begin{eqnarray}
	(E^{(0)}_{\alpha,{\bf k}})^2 = \frac{\omega_0}{\{(\omega_0+a_\alpha({\bf k}))^2-b^2({\bf k})\sin^22\theta\}\{\omega_0+a_\alpha({\bf k})\}},\nonumber\\
\end{eqnarray}
and
\begin{eqnarray}
f_n(x) = \frac{(2n-3)!!}{2^{n-1}n!} x^{n-1}.
\end{eqnarray}
As $E^{(0)}_{\alpha,{\bf k}}>0$ and $f_n(x)\ge0$ for arbitrary $n=1,3,5,\cdots$, it is shown that $E_{\alpha,{\bf k}}\ge0$ when $J_3\ne0$. Hence, the sign of the coefficient for $\sin4\theta$ in Eq.~(\ref{eq:dE_dth}) is
\begin{eqnarray}
-\frac14\sum_{\alpha,{\bf k}} E_{\alpha,{\bf k}}\,b^2({\bf k}) \le 0
\end{eqnarray}
This indicates that $E_\text{GS}$ has a minimum at $\theta=\frac{2\pi n + 1}4$. Hence, whenever $J_3\ne0$, the quantum fluctuations give rise to four-fold magnetic anisotropy favoring the $\langle110\rangle$ directions, i.e. with the same sign as the effect of thermal-fluctuations. As quantum fluctuations dominate as $T\to 0$, it is expected that they will modify the $H_{c1}$ in Fig.~\ref{fig:mc_diag} to remain non-zero at $T\to0$.

\section{Two-sublattice Model} \label{sec:resB}

In this section, we study the magnetic behavior of the two-sublattice model in Sec.~\ref{sec:stagg-tetr-case}. For simplicity, in this section, we mainly consider the Hamiltonian in Eq.~(\ref{eq:model:H2S}). In Sec.~\ref{ssec:resB:gs}, we study the classical ground state properties of the model. By using a variational method, we show that a wide range of the phase diagram is dominated by canted-ferromagnetic order. The thermodynamics of the canted-ferromagnetic state is studied in Sec.~\ref{ssec:resB:mc} using classical Monte-Carlo simulation. 

\subsection{Ground state}\label{ssec:resB:gs}

\begin{figure}
\includegraphics[width=.8\linewidth]{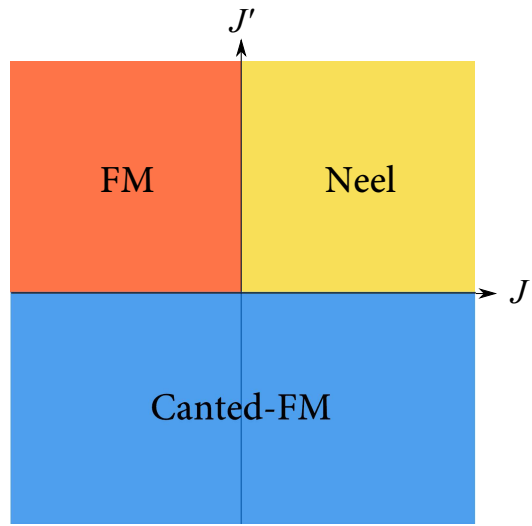}
\caption{
Phase diagram of Eq.~(\ref{eq:model:H2S}) obtained by the two sublattice variational calculation. Here FM (Neel) denotes a ferromagnetic (antiferromagnetic) state with spins pointing parallel to $z$ axis. The spins point in the $xy$ plane in the canted ferromagnetic (Canted-FM) state.
}\label{fig:gstwosub}
\end{figure}

We first present the ground state phase diagram of the model in Eq.~(\ref{eq:model:H2S}).  To make progress, we assume a two-sublattice structure, and find the classical ground state within this space.  This is equivalent to finding the best variational product state for the S=1/2 model with a two-sublattice structure.  The resulting phase diagram is shown in Fig.~\ref{fig:gstwosub}.

\subsubsection{Zero magnetic field}
\label{sec:zero-magnetic-field}

We observe three different phases. When $J<0$ and $J^\prime>0$, spins align ferromagnetically along the $z$ axis in spin space. On the other hand, when $J>0$ and $J^\prime>0$, we obtain antiferromagnetic order with spins aligned along $z$ axis, antiparallel on the two sublattices.   We note that, in this region of the phase diagram, an arbitrary spin state which satisfies the two-up two-down local constraint on each tetrahedron is a ground state. Hence, the ground state retains quasi-macroscopic degeneracy, at the minimum.  This degeneracy is of course unstable to additional interactions.

On the other hand, if $J^\prime<0$, the phase diagram is dominated by a canted ferromagnetic state. This is a two-sublattice magnetic order in which the spins on different sublattices are non-collinear (see Fig.~\ref{fig:intro:fcc}). Spins lie within the $xy$ plane with angles given by
\begin{eqnarray}
\tan\phi_A &=& -\frac{J^\prime}{4|J|}-\sqrt{1+\left(\frac{J^\prime}{4J}\right)^2}\\
\phi_B     &=& \frac{\pi}2-\phi_A
\end{eqnarray}
for $J>0$ and
\begin{eqnarray}
\tan\phi_A &=& \frac{J^\prime}{4|J|}+\sqrt{1+\left(\frac{J^\prime}{4J}\right)^2}\\
\phi_B     &=& \frac{\pi}2-\phi_A
\end{eqnarray}
for $J<0$. Here, 
\begin{eqnarray}
{\bf S}_\alpha = (\cos\phi_\alpha,\sin\phi_\alpha,0)
\end{eqnarray}
with $\alpha=A,B$.   Adding the two non-collinear moments, we see that this phase sustains a net uniform magnetic moment along the $\langle110\rangle$ direction.  The difference of the two moments gives the staggered magnetization perpendicular to the uniform magnetic moment. The ground state has a four-fold degeneracy, comprised of state related by time-reversal symmetry and a mirror reflection through the $\left\{100\right\}$ plane.   We note that this state is similar to one found in a previous study on a $J_\text{eff}=3/2$ model.~\cite{Chen2010}

\begin{figure}
\includegraphics[width=.9\linewidth]{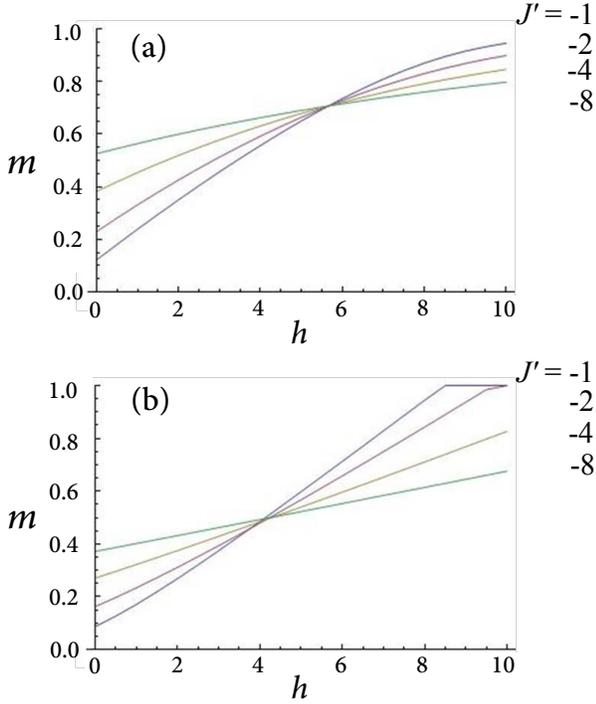}
\caption{
Magnetization process of the canted FM state with field applied along (a) $[110]$ and (b) $[100]$ direction.
$J=1$ is used as the unit of energy.
}\label{fig:cFM_magproc}
\end{figure}

\subsubsection{Non-zero applied field}
\label{sec:non-zero-applied}

When an external magnetic field is applied, the canted ferromagnetic phase evolves with characteristic magnetization curves that depend upon the direction of the external field. Fig.~\ref{fig:cFM_magproc}(a) shows the field dependence of uniform magnetic moment along $[110]$ direction for different values of $J^\prime$. When $h=0$, one finds a finite magnetic moment that is dependent on $J^\prime/J$. When the magnetic field applied along $[110]$, the magnetic moment increases monotonically without and sharp features, approaching $m\to1$ at $h\to \infty$. This is a consequence of the $J^\prime$ interaction which induces moment canting for arbitrarily small $J'/h$.  To see this, we consider the $h\gg J, J^\prime$ limit, in which the leading correction to the energy from $h$ appears in the form
\begin{eqnarray}
E &\sim& 6J+2J^\prime - 2h + \frac{h}4 (\delta \phi_+^2 + \delta \phi_-^2)\nonumber\\
&&\quad - 4J^\prime \delta \phi_- - 2J \delta \phi_-^2 + O(\delta\phi_\alpha^3).\label{eq:cFM_h110}
\end{eqnarray}
Here, 
\begin{eqnarray}
\delta\phi_\pm = \delta\phi_A \pm \delta\phi_B
\end{eqnarray}
with
\begin{eqnarray}
\delta\phi_\alpha = \phi_\alpha - \frac\pi4
\end{eqnarray}
and $h$ is the external field applied along $[110]$ axis,
\begin{eqnarray}
{\bf h} = \frac{h}{\sqrt2}(1,1,0).
\end{eqnarray}
The presence of the linear $J' \delta\phi_-$ term in Eq.~(\ref{eq:cFM_h110}) shows that the fully polarized state at $h = \infty$ is unstable to infinitesimally small $J^\prime$.

By contrast, for a field applied along the $[100]$ axis, a sharp saturation feature appears, as shown in Fig.~\ref{fig:cFM_magproc}(b). The figure shows $[100]$ component of the net magnetic moment. We see that the magnetization reaches its saturation value at
\begin{eqnarray}
\frac{h}4=J+\sqrt{J^2+\left(\frac{J^\prime}2\right)^2}.
\end{eqnarray}
This feature is a characteristic which can distinguish the canted FM state.

\subsection{Classical Monte Carlo Simulation}\label{ssec:resB:mc}

\begin{figure*}
\includegraphics[width=\linewidth]{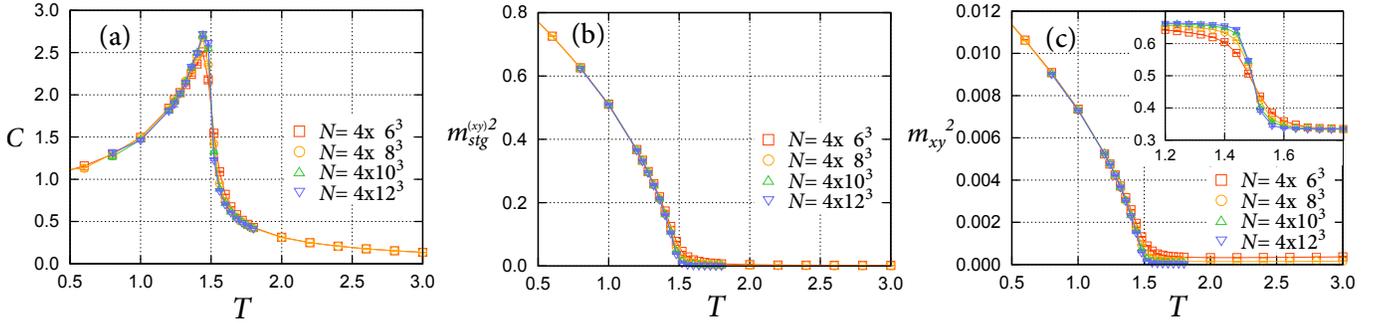}
\caption{
Temperature dependence of (a) specific heat and square of the $xy$ components of (b) staggered magnetization $m^{(xy)}_{\rm stg}{}^2$ and (c) uniform magnetization $m_{xy}^2$. The calculation were done under zero external field at $J=1$ and $J^\prime=-1$. Inset in (c) shows Binder parameter for $m_{xy}^2$.
}\label{fig:cFM_mc1}
\end{figure*}

The presence of the canted-ferromagnetic phase is confirmed by the Monte Carlo simulation. Fig.~\ref{fig:cFM_mc1} shows the temperature dependence of specific heat and magnetic structure calculated by classical Monte Carlo simulation. The temperature dependence of specific heat is shown in Fig.~\ref{fig:cFM_mc1}(a). It shows a peak at $T\sim1.5$ indicating a phase transition which is associated with the rapid increase of $m^{(xy)}{}^2$ and $m_\text{stg}^{(xy)}{}^2$ [shown in Fig.~\ref{fig:cFM_mc1}(b) and (c), respectively]. The square of the $xy$ compoment of net magnetization $m^{(xy)}{}^2$ is defined in the same way as $m_\text{stg}^{(xy)}{}^2$ [see Eq.~(\ref{eq:mstg_xy})]. The results indicate a transition to the canted-FM state. The transition temperature $T_c^{(cFM)}=1.49(3)$ was estimated from the Binder parameter for $m^{(xy)}$, which is shown in the inset of Fig.~\ref{fig:cFM_mc1}(c).

\begin{figure*}
\includegraphics[width=\linewidth]{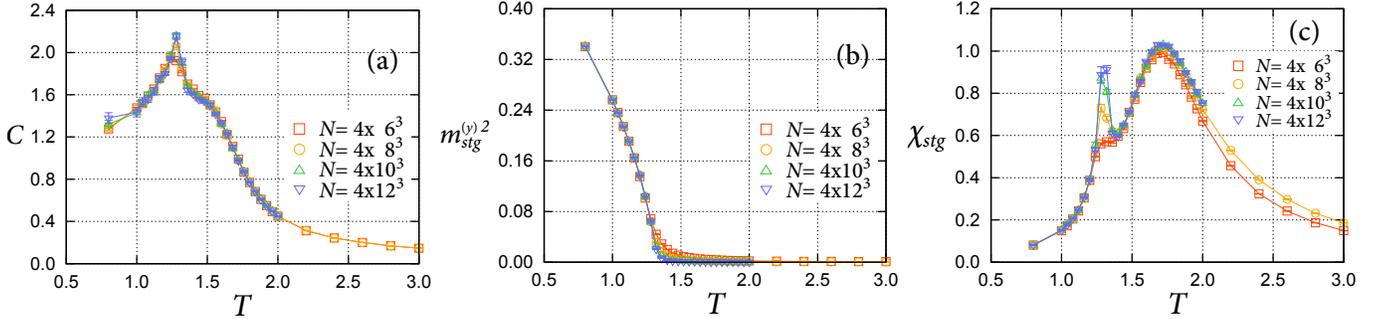}
\caption{
Temperature dependence of (a) specific heat, (b) $y$ component of staggered magnetization, and (c) the susceptibility of staggered magnetization. The calculation were done at $J=1$, $J^\prime=-1$, and $h=1.2$.
}\label{fig:cFM_mc2}
\end{figure*}

With the application of an external field along $[100]$ axis, the four-fold degeneracy of the ground state is lifted. However, a two-fold degeneracy remains, related to mirror symmetry in the [100] plane. Hence, at low temperature, the system is expected to show a phase transition associated with spontaneous symmetry breaking that selects one of the two ground states. This is observed as a peak at $T\sim1.3$ in the specific heat shown in Fig.~\ref{fig:cFM_mc2}(a). Associated with this peak, the $y$ component of the staggered magnetization shown in Fig.~\ref{fig:cFM_mc2}(b) shows a rapid increase.

In addition to this phase transition, in the MC simulation in Fig.~\ref{fig:cFM_mc2}, a crossover from the high-$T$ paramagnetic state to a low-$T$ FM-like paramagnetic state was observed.  This is not a phase transition but can be regarded as a rapid crossover related to the zero field transition. In Fig.~\ref{fig:cFM_mc2}(a), the specific heat shows a shoulder around $T\sim1.5$. In addition, a hump is observed in the Monte Carlo simulation of the susceptibility of the staggered moment [see Fig.~\ref{fig:cFM_mc2}(c)].

\begin{figure}
\includegraphics[width=.9\linewidth]{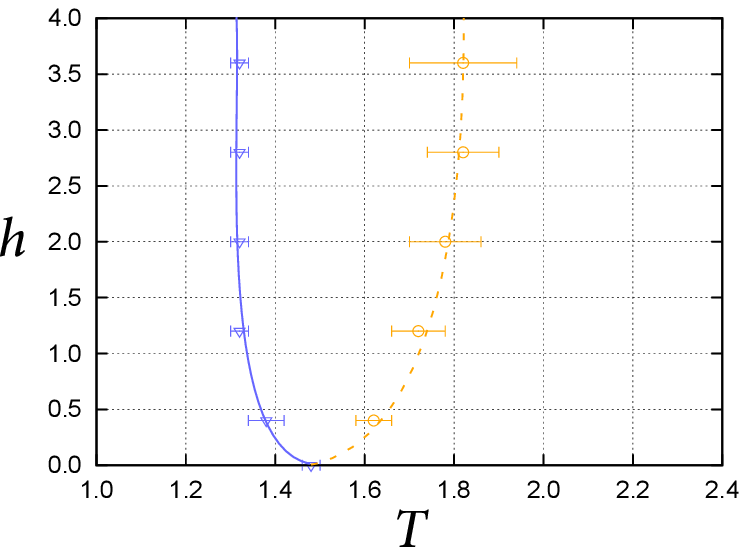}
\caption{
$T$-$H$ phase diagram in the classical limit for the Hamiltonian in Eq.~(\ref{eq:model:H2S}). The magnetic field is applied along the $[100]$ direction. The solid circles indicate the phase boundary between the paramagnetic phase and canted-ferromagnetic state; the boundary was estimated from the Binder parameter. The solid and open squares indicate the crossover temperature from the high-$T$ paramagnetic state to intermediate paramagnetic state. Solid and open symbols are estimated from the maximum of the hump in specific heat and susceptibility of the staggered moment.
}\label{fig:cFM_pd}
\end{figure}

The critical and the crossover temperatures estimated from the specific heat and the hump of the susceptibility are plotted in Fig.~\ref{fig:cFM_pd}. At $h=0$, both the transitions take place at the same temperature: namely, a single transition to the canted-FM state is observed. With increasing $h$, the temperature of onset of $m$ and $m_\text{stg}^{(y)}$ separate. Under ambient magnetic field, the transition temperature to the canted-FM state shows a slight decrease around $h\to0$, but is nearly independent of the applied field through out the range of calculation. On the other hand, the crossover temperature is enhanced by $h$.

\section{Spin-Phonon Coupling}\label{ssec:resB:gl}

In this section, we consider the coupling of spins to lattices. In particular, we investigate the instability of lattices with tetragonal symmetry, assuming the spins to be of XY type with $C_4$ symmetry. In Sec.~\ref{ssec:resB:gl:formula}, we introduce the Ginzburg-Landau (GL) theory we consider throughout this section. The effect of spin-lattice coupling in the weakly-coupled regime is studied in Sec.~\ref{sec:weak-spin-lattice} and a mean-field argument for stronger coupling is presented in Sec.~\ref{sec:elastic-mean-field}. The expected phase diagram based on these arguments are described in Sec.~\ref{sec:phase-transitions}.

\subsection{Ginzburg-Landau Theory}
\label{ssec:resB:gl:formula}

To investigate the effect of spin-lattice coupling, we begin by writing an effective GL free energy as the sum of all three terms,
\begin{eqnarray}
F=F_{\phi} + F_{u} + F_{u\phi}.
\end{eqnarray}
The spin part is given by
\begin{eqnarray}
F_{\phi}&=&\int d^d x\; \frac12 t |\phi|^2 + \frac1{2m_z}\partial_z\bar{\phi}\partial_z\phi + \frac1{2m_1}\partial_\alpha\bar{\phi}\partial_\alpha\phi \nonumber\\
&&+ \frac1{2m_2}\{(\partial_x\phi)^2-(\partial_y\phi)^2+\text{h.c.}\}\nonumber\\
&&+ \frac{\mathrm{i}}{2m_4}(\partial_x\phi\partial_y\phi - \text{h.c.})-\frac12 (h\bar\phi + \text{h.c.}) \nonumber\\
&&+ u |\phi|^4+ v(\phi^4+\bar{\phi}^4),\label{eq:Fphi}
\end{eqnarray}
where $\phi=\phi({\bf r})=\phi_x({\bf r})+\mathrm{i}\phi_y({\bf r})$ is a complex order parameter field for the spins composed of the two components of the spins $\phi_{\alpha}({\bf r})$ ($\alpha=x,y$), and $\bar{\phi}=\bar{\phi}({\bf x})$ is the complex conjugate of $\phi$.
\begin{eqnarray}
h = h_x + \mathrm{i}h_y
\end{eqnarray}
is the complex representation of the external magnetic field with $h_\alpha$ being the external field along $\alpha$ axis. The magnetic transition of $F_\phi$ is tuned by the reduced temperature $t \sim (T-T_{c;MF})/T_{c;MF}$, where $T_{c;MF}$ is the mean field approximation to $T_c$. 

The lattice free energy describes long wavelength phonons, corresponding to the uniform distortion of the lattice. The general form of the elastic free energy allowed by tetragonal symmetry is 
\begin{eqnarray}
\label{eq:S_lattice}
F_u & = & \int d^3x \; f_u(\epsilon_{ab}), \\
f_u(\epsilon_{ab}) & = & \frac12 c_{11} (\epsilon_{xx}^2+ \epsilon_{yy}^2) + \frac12 c_{33}\, \epsilon_{zz}^2 + c_{12}\, \epsilon_{xx} \epsilon_{yy}\nonumber \\
&& + c_{13}(\epsilon_{xx}\epsilon_{zz}+\epsilon_{yy}\epsilon_{zz}) + 2 c_{44}(\epsilon_{yz}^2+\epsilon_{xz}^2) \nonumber \\
&& + 2c_{66}\, \epsilon_{xy}^2,
\end{eqnarray}
with $\epsilon_{\alpha\beta}=\epsilon_{\alpha\beta}({\bf r})$ ($\alpha,\beta=x,y,z$) the strain tensor
\begin{eqnarray}
\epsilon_{\alpha\beta} = \frac12(\partial_\beta u_\alpha + \partial_\alpha u_\beta),
\end{eqnarray}
and $u_\alpha=u_\alpha({\bf r})$ is the phonon mode that corresponds to lattice distortion along $\alpha=x,y$. For simplicity, we here assumed the lattice is rigid against distortion along $z$ axis.  The constants $c_{ab}$ follow standard definitions.  Structural transitions are described within elastic theory by vanishing eigenvalues of the elastic tensor, and have been discussed for example by Cowley.~\cite{Cowley1976}  In particular, tetragonal to orthorhombic transitions, which may be expected based on spin physics, arise when either $c_{11}-c_{12}$ or $c_{66}$ vanishes.  The other elastic coefficients describe shears and other possible deformations of less interest here.

By symmetry, the coupling of the spins and phonons is given by
\begin{eqnarray}
F_{u\phi}&=&\int d^d x\; \frac12\left\{ \lambda_1 (\epsilon_{xx} + \epsilon_{yy})+\lambda_2\epsilon_{zz}\right\} |\phi|^2 \nonumber\\
&&\qquad+ \lambda_{x^2-y^2}(\epsilon_{xx} - \epsilon_{yy})\,{\rm Re}(\phi^2) \nonumber\\
&&\qquad+ \lambda_{xy} \epsilon_{xy}\,{\rm Im}(\phi^2). \label{eq:sph_comp}
\end{eqnarray}
It may be convenient to define 
\begin{equation}
  \label{eq:2}
  \epsilon = \frac{1}\lambda \left[\lambda_{x^2-y^2} (\epsilon_{xx}-\epsilon_{yy}) + i \lambda_{xy} \epsilon_{xy}\right],
\end{equation}
with
\begin{equation}
  \label{eq:4}
  \lambda = \sqrt{\lambda_{x^2-y^2}^2 + \lambda_{xy}^2},
\end{equation}
and $\bar\epsilon = \epsilon^*$.  Then this can be rewritten as
\begin{eqnarray}
  \label{eq:3}
  F_{u\phi}&=&\int d^d x\; \frac12\left\{ \lambda_1 (\epsilon_{xx} + \epsilon_{yy})+\lambda_2\epsilon_{zz}\right\} |\phi|^2 \nonumber\\
&&\qquad+ \frac{\lambda}{2}\left( \bar\epsilon \, \phi^2 + \epsilon \bar\phi^2\right).
\end{eqnarray}

\subsection{Weak spin-lattice coupling}
\label{sec:weak-spin-lattice}

One may consider the perturbative regime of weak spin-lattice coupling.  For zero coupling, $\lambda_a=0$, the spin and lattice sub-systems are independent.  If we assume stability of the disconnected lattice subsystem, then there are three regimes to consider, dictated by the state of the spins: $T>T_c$, $T \approx T_c$, and $T<T_c$. 

For $T>T_c$, the spins are disordered, and $\phi$ undergoes small fluctuations about $\phi=0$.  The full tetragonal symmetry is maintained, and perturbation theory in $\lambda_a$ will lead only to small corrections to the elastic theory.  

For $T<T_c$, the spin system on its own breaks the tetragonal symmetry.  Depending upon the sign of $v$, states with $\langle \phi^2\rangle$ non-zero and real ($v<0$) or imaginary ($v>0$) are selected.  This corresponds to magnetic order aligned along $\langle 100\rangle$ or $\langle 110\rangle$ axis, respectively.  This non-zero expectation value reduces $F_{u\phi}$ in Eq.~\eqref{eq:sph_comp} at first order to a linear term in the strains, which induces non-zero $\epsilon_{xx}-\epsilon_{yy}$ (for $v<0$) or $\epsilon_{xy}$ (for $v>0$).  Hence the lattice responds to the symmetry breaking dictated by the spins.

For $T \approx T_c$, na\"ive perturbation theory is invalidated by the power-law correlations of the criticality of the spins.  Instead, we can analyze the stability of the critical point using renormalization group arguments.  The decoupling spin system described by Eq.~\eqref{eq:Fphi} in zero magnetic field has a $Z_4$ symmetry under $\pi/2$ rotations of $\phi$.  It is established that the fourfold anisotropy ($v$) is irrelevant at the critical point in three dimensions, so that the fourfold symmetry is enlarged to continuous rotational invariance in the $xy$ plane.  The universal properties of the system at the critical point are described by the scale invariant O(2) Wilson-Fisher fixed point field theory.  This field theory contains two independent composite operators quadratic in the $xy$ field.  The first is the ``energy density'' operator $|\phi|^2$, which has scaling dimension $[|\phi|^2]=d-1/\nu$, where $d=3$ is the dimensionality and $\nu \gtrsim 2/3$ is the correlation length exponent (the fact that $\nu>2/3$ follows from hyperscaling, which requires $\nu= (2-\alpha)/3$, where $\alpha$ is the specific heat exponent, which is known to be slightly negative for the 3d XY model).  Hence $[|\phi|^2] \gtrsim 3/2$.  The second operator is the symmetric tensor field $\phi^2$, which because it is affected by the fluctuations of the phase of $\phi$, is expected intuitively to have a larger scaling dimension than the energy density.  An estimate from the $\varepsilon$-expansion~\cite{Oshikawa2000} gives $[\phi^2] \approx 9/5$.  The scaling dimension of the strain is determined by the requirement that the total free energy be dimensionless, hence $[\epsilon_{ab}]=3/2$ for all strain components in $d=3$.  

The full scaling dimension of the perturbations in Eq.~\eqref{eq:sph_comp} is $[\epsilon_{ab}]+[|\phi|^2]$ and $[\epsilon_{ab}]+[\phi^2]$.  Using the above results, we see that all these total scaling dimensions are {\it larger} than $3$, which implies that the terms in $F_{u\phi}$ are {\it irrelevant} at the XY fixed point.  This demonstrates that spin-lattice coupling effects are {\em irrelevant} when weak, and a single direct transition in the XY universality class is maintained.  

\subsection{Elastic mean field}
\label{sec:elastic-mean-field}

By inspection, the terms in $F_{u\phi}$ can lower the free energy of orthorhombically distorted states independent of whether they exhibit magnetic order.  Hence,  despite the fact that, when they are weak spin-lattice coupling does not induce a nematic phase, we expect the nematic will occur when the appropriate $\lambda_a$ is sufficiently large.  To go beyond the perturbative approach, we consider a mean-field treatment of the elasticity, while continuing to include all fluctuation effects in the spins.  Formally, we define the effective free energy for $\epsilon_{ab}$ ($u_\alpha$) obtained by integrating out $\phi$ completely: 
\begin{equation}
  \label{eq:1}
  F_{\rm eff}[\epsilon_{ab}] = F_u - \log \left[ \int [d\phi] \exp\left ( - F_\phi - F_{u\phi}\right) \right],
\end{equation}
where $\int [d\phi]$ indicates the full functional integral over $\phi$.  We treat $F_{\rm eff}[\epsilon]$ in mean-field, i.e. just seeks its thermodynamic minimum.  The coefficients $\lambda_1,\lambda_2$ play no essential role, as they do not couple symmetry breaking order parameters (they instead just describe how changes in the lattice parameters of the tetragonal structure are coupled to the energy density of spins), and so they can be set to zero.   

In mean field, we assume that the tensor $\epsilon_{ab}$ is a constant.   Then the free energy is simply the system volume times the free energy density,
\begin{equation}
  \label{eq:5}
  f_{\rm eff}(\epsilon_{ab}) = f_u(\epsilon_{ab}) + \delta f(\epsilon),
\end{equation}
where in general, to consider the regime of non-zero orthorhombic distortions, we should augment $f_u$ by third and fourth order terms in $\epsilon_{ab}$ for stability. 

If we focus on the case of zero field $h=0$ and assume we are above the order temperature for the magnetism, $T>T_c$, then all the terms which are irrelevant at the magnetic critical point -- $v$, $m_2^{-1}$, and $m_4^{-1}$ -- are unimportant and can be neglected.  Then  the isotropic gradient coefficient can be set to unity by rescaling coordinates: $m_z, m_1 \rightarrow 1$.  Thus finally, we see that $\delta f$ is a function of $t$, $u$, and $\lambda\epsilon$.  Using scaling around the Gaussian fixed point where $t=u=\lambda=0$, we have (using $d=3$)
\begin{equation}
  \label{eq:6}
  \delta f =  -b^{-3} \mathcal{F}( b^2 \,t, b u, b^2 \lambda |\epsilon|).
\end{equation}
The free energy can depend only upon the absolute value of $\epsilon$ by U(1) symmetry.
By choosing $b = t^{-1/2}$, we then obtain
\begin{equation}
  \label{eq:7}
  \delta f = -t^{3/2} \mathcal{F}(\frac{u}{\sqrt{t}}, \frac{\lambda |\epsilon|}{t}),
\end{equation}
with $\mathcal{F}(x,y) \equiv \mathcal{F}(1,x,y)$.  We used an overall minus sign in the definitions of the scaling functions in Eqs.~(\ref{eq:6}) and (\ref{eq:7}), anticipating that $\delta f$ is negative.

\subsubsection{Structural instability point}
\label{sec:struct-inst-point}

To check the {\sl local} stability of the tetragonal state, we expand the free energy to quadratic order in $\epsilon$ and check whether the resulting quadratic form is stable.  We have
\begin{equation}
  \label{eq:9}
  f_{\rm eff}^{(2)} = f_u(\epsilon_{ab}) - \frac{\lambda^2}{2\sqrt{t}} \mathcal{F}_{0,2}(\frac{u}{\sqrt{t}},0) |\epsilon|^2.
\end{equation}
The latter term may be absorbed into renormalized elastic moduli according to 
\begin{subequations} \label{eq:10}
\begin{eqnarray}
  c_{11} & \rightarrow&  \tilde{c}_{11} = c_{11} - \frac{\lambda_{x^2-y^2}^2}{\sqrt{t}} \mathcal{F}_{0,2}(\frac{u}{\sqrt{t}},0), \\
  c_{12} & \rightarrow & \tilde{c}_{12} = c_{12} + \frac{\lambda_{x^2-y^2}^2}{\sqrt{t}} \mathcal{F}_{0,2}(\frac{u}{\sqrt{t}},0),  \\
  c_{66} & \rightarrow & \tilde{c}_{66} = c_{66} - \frac{\lambda_{xy}^2}{4\sqrt{t}} \mathcal{F}_{0,2}(\frac{u}{\sqrt{t}},0). 
\end{eqnarray}
\end{subequations}
We see that indeed $\tilde{c}_{11}-\tilde{c}_{12}$ and $\tilde{c}_{66}$ are reduced by the spin fluctuations, and if their bare values are small enough, the renormalized values become negative, signaling an elastic instability.  

\subsubsection{Gaussian regime}
\label{sec:gaussian-regime}

We may consider two regimes.  First, when $t \gg u^2$, the first argument is small, and one can neglect the quartic interaction term.  This is the Gaussian regime.  The inequality $t \gg u^2$ corresponds to the Ginzburg criterion, under which mean field behavior (for $\phi$) is expected.   In the above calculation of the instability point, one may take $\mathcal{F}_{0,2}$ as constant in this regime, and we observe that the corrections to the elastic modulo become arbitrarily large for small enough $t$.  This seems in conflict with the conclusion of the previous subsection that weak spin-lattice coupling does not destabilize the XY transition.  We will return to this point.

Beyond just the instability analysis, in this case, the full free energy may be explicitly calculated, since the functional integral in Eq.~\eqref{eq:1} is quadratic.   One finds simply
\begin{equation}
  \label{eq:8}
  \mathcal{F}(0,y) = -\frac{1}{4\pi^2} \int_0^\infty dk k^2 \ln \left[ \frac{(1+k^2)^2 - y^2}{(1+k^2)^2}\right],
\end{equation}
which is well-defined only for $|y|<1$.  By Taylor expansion, one obtains to fourth order
\begin{eqnarray}
  \label{eq:11}
  \mathcal{F}(0,y) = \frac{\pi}{64} y^2 (16+ y^2) + \mathcal{O}(y^6).
\end{eqnarray}
This determines the instability points depending upon the values of $c_{11}-c_{12}$, $c_{66}$, and $\lambda_a$, according to the condition $\tilde{c}_{11}-\tilde{c}_{12}=0$ or $\tilde{c}_{66}=0$.  Looking further, however, we see that the contribution from spin fluctuations to the fourth order term in the free energy as a function of $|\epsilon|$ is {\it negative}.  This is suggestive of a first order transition.  To explore this further, let us assume for concreteness an instability in the $\epsilon_{xx}-\epsilon_{yy}$ channel.  We take $\lambda_{xy}=0$ and $\epsilon_{xx} = -\epsilon_{yy}= - \epsilon/2$ and all other components of the strain tensor zero.  Then the elastic free energy is
\begin{eqnarray}
  \label{eq:12}
  f_u(\epsilon) = \frac{c_{11}-c_{12}}{4} \epsilon^2 + w \epsilon^4 + \mathcal{O}(\epsilon^6),
\end{eqnarray}
where we included a fourth order term for stability.  Including the second and fourth order terms from $\delta f$, we obtain then
\begin{equation}
  \label{eq:13}
  f_{\rm eff} = \frac{\tilde{c}_{11}-\tilde{c}_{12}}{4} \epsilon^2 + (w - \frac{\pi \lambda^4}{64 t^{5/2}})\epsilon^4 + \mathcal{O}(\epsilon^6).
\end{equation}
Suppose $\lambda$ is small, and $c_{11}-c_{12}$ is fixed.  Then in the Gaussian regime, we see that the elastic instability occurs [see Eqs.~\eqref{eq:10}] at the critical reduced temperature $t_{\rm el} \sim (\frac{\lambda^2}{c_{11}-c_{12}})^2$.  For this value $t=t_{el}$ the negative term in the coefficient of $\epsilon^4$ in Eq.~\eqref{eq:13} scales as $(c_{11}-c_{12})^5/\lambda^6$, which is much larger than the constant $w$ for small $\lambda$.  Hence in this limit it seems that the fourth order term in the free energy $f_{\rm eff}$ is negative, and indeed a first order elastic transition obtains.

\subsubsection{Universal $xy$ regime}
\label{sec:universal-xy-regime}

However, this conclusion rests on the Gaussian treatment of the $\phi$ field.  This breaks down close to the XY critical point, when $u/\sqrt{t}$ is no longer small.  When $u/\sqrt{t}$ becomes large, we expect that the universal critical exponents of the XY Wilson-Fisher fixed point should apply.  Therefore in this limit we can deduce that
\begin{equation}
  \label{eq:14}
  \mathcal{F}(x,y) \sim x^{-6\nu +3} \mathcal{F}(x^{-(2+ ([\phi^2]-3)\nu)} y),
\end{equation}
where $\mathcal{F}(z)$ is a new universal function.  This is required so that the proper universal scaling behavior
\begin{equation}
  \label{eq:15}
  \delta f \sim - t^{3\nu} \mathcal{F}( \lambda t^{([\phi^2]-3)\nu} |\epsilon|)
\end{equation}
obtains.  By expansion, we now see that in the XY regime, the renormalized elastic modulus takes the form
\begin{equation}
  \label{eq:16}
  \tilde{c}_{11}-\tilde{c}_{12} = c_{11}-c_{12} - \lambda^2 t^{(2[\phi^2]-3)\nu},
\end{equation}
where $(2[\phi^2]-3)\nu \approx 2/5$ is positive.  We see that, contrary to the Gaussian regime, the correction to the elastic modulus {\it decreases} with reducing $t$ on approaching close to the critical point.  Note furthermore that for small $t$ the XY regime is always entered, cutting off the apparent Gaussian divergence.  This means that for sufficiently small $\lambda$, where the instability criterion is not reached before the crossover to the XY regime, there is no instability and the XY critical point is stable, as claimed in the prior subsection.

\subsubsection{Crossover regime}
\label{sec:crossover-regime}

The above analysis implies that the renormalization of the elastic modulus for small $\lambda$ is non-monotonic with temperature.  According to scaling, for example
\begin{equation}
  \label{eq:17}
  \tilde{c}_{11}-\tilde{c}_{12} = c_{11}-c_{12} - \frac{\lambda^2}{\sqrt{t}} \mathcal{C}(\frac{u}{\sqrt{t}}),
\end{equation}
where $\mathcal{C}(0)$ is a constant, and $\mathcal{C}(x) \sim x^{-9/5}$ for $x \gg 1$.  Hence the (negative) renormalization of the elastic modulus is maximum for $t\sim u^2$, and is of order $\lambda^2/u$.  This sets a threshold value for an instability at $\lambda = \lambda_{\rm min} \sim \sqrt{u(c_{11}-c_{12})}$.  For $\lambda>\lambda_{\rm min}$, a nematic phase appears in an interval of temperature around $t \sim u^2$.  Remarkably, the elastic mean field theory predicts at for $\lambda$ close to $\lambda_{\rm min}$, the nematic phase exists as an ``island'' {\it within} the tetragonal phase, i.e. there is a re-entrant tetragonal phase at {\it lower} temperature than the nematic one, before the XY transition occurs to the orthorhombic magnetic state.  

With further increase of $\lambda$, the extent of nematic phase grows until its lower boundary reaches the XY critical point.  For larger $\lambda$, the re-entrant tetragonal phase no longer exists, and the system evolves through the simpler sequence of tetragonal to nematic (orthrhombic) to magnetic phases on reducing temperature.  

\subsection{Phase transitions}
\label{sec:phase-transitions}

The above analysis predicts several different phase transitions when spin-lattice coupling is substantial.  A tetragonal to orthorhombic (which may also be called tetragonal to nematic) transition may occur {\it twice} at intermediate $\lambda$, or once at larger $\lambda$.  The elastic mean field theory predicts the higher temperature transition is first order, while the lower transition may be second order.  A continuous second order transition is in fact known to be possible based on renormalization group analysis,~\cite{Cowley1976} and if this transition is continuous it is expected to display mean-field critical exponents owing to the influence of long-range elastic forces.  The lower transition from nematic/orthorhombic to the magnetic phase is of Ising type.  Finally, the phase diagram contains an interesting multicritical point where the nematic, magnetic, and tetragonal phases meet.  To our knowledge, the universality class of this multicritical point has not been studied carefully, and is left as an interesting problem for future theory.

\section{Summary and Discussion}
\label{sec:discussions-summary}

In this paper, we studied the magnetic properties of two spin models on a fcc lattice with tetragonal lattice symmetry, using several analytical techniques and Monte Carlo simulations.  The results should be applicable to insulating magnetic double perovskites with a single magnetic species, in which orbital degeneracy is broken.  

In the first half of the study, we considered the simplest tetragonal model corresponding to a uniform lattice distortion of the cubic fcc system along $z$ axis. We mapped out the ground state phase diagram in the classical limit, and showed that it is dominated by four magnetic phases, $xy$FM, $xy$AFM, $z$FM, and $z$AFM orders.   The first two of these phases were shown to exhibit the phenomena of ``order by disorder'': accidental classical degeneracy lifted by fluctuations.  As a result, magnetic polarization along the $\langle110\rangle$ directions was favored.

In the latter half of the paper, we considered a more complex tetragonal model in which the tetragonal symmetry is realized through two inequivalent fcc sites, related by a screw axis. This is suggested by earlier theory as a result of orbital/quadrupolar order.\cite{Chen2010}\ After writing the general Hamiltonian for this system, we focused upon a physically motivated parameter regime, and showed that the model exhibits a peculiar canted ferromagnetic state. 

In addition, we studied the effect of spin-lattice coupling, and pointed out the possibility of a nematic transition induced by the coupling. It might be interesting to measure a ``nematic susceptibility'', as has been done by elasto-resistivity\cite{Chen2012}.  For an insulator, the resistivity anisotropy may be too difficult to measure.  Hence we suggest this could be done instead by identifying the electronic nematic order parameter $\psi$  with the anisotropy of the magnetic susceptibility, i.e.
\begin{equation}
  \label{eq:18}
  \psi = \frac{\chi_{xx}-\chi_{yy}}{\chi_{xx}+\chi_{yy}}.
\end{equation}
Then the nematic susceptibility may be defined as $\chi_n = \partial \psi/\partial \epsilon$ (here we consider the nematic order corresponding to a $\langle$100$\rangle$ deformation $\epsilon = \epsilon_{xx}-\epsilon_{yy}$, but one can make a similar definition for a $\langle$110$\rangle$ deformation $\epsilon = \epsilon_{xy}$ by a 45 degree rotation).  In the phase diagrams of Fig.~\ref{fig:intro:phdiag}, we expect behavior of $\chi_n$ appropriate to a composite nematic order parameter $\psi \sim {\rm Re}\, \phi^2$ on the lower boundary, and of a fundamental Ising-nematic order parameter on the upper right boundary.  

Our theory presented here is potentially applicable to wide range of materials in the double-perovskite family.~\cite{Chen2010}  It complements an earlier study,\cite{Chen2010}\ going into more depth with fewer theoretical assumptions, for the case of tetragonal symmetry.   The most obvious application is to Ba$_2$NaOsO$_6$, which exhibits a ferromagnetic ground state with an unusual $\langle 110\rangle$ easy axis.~\cite{Erickson2007}   This easy axis is readily explained if the cubic symmetry is broken to tetragonal, and we assume here that this occurs at a temperature high compared to the magnetic ordering.  Evidence of this appears to have been found recently by x-ray scattering.\cite{IslamUP}\  Further comparison with structural data from x-rays at low temperature,\cite{IslamUP} and with NMR and NQR measurements that may discern details of the magnetic and tetragonal ordering,\cite{MitrovicUP} should be fruitful.  

More generally, we would like to emphasize that the combination of strong spin-orbit coupling, narrow electronic bandwidth, and varieties of structural motifs of double perovskites makes them potentially a rich realization of highly quantum frustrated spin-1/2 Hamiltonians with exotic directional-dependent spin couplings.  In this, they comprise another family to complement the honeycomb iridate family which has been much studied recently.

\acknowledgements

The authors thank I. R. Fisher, Z. Islam, V. Mitrovic, and E.-G. Moon for fruitful discussions. Part of the calculations were done at Center for Scientific Computing at California Nanosystems Institute and Material Research Laboratory, University of California Santa Barbara: an NSF MRSEC (DMR-1121053) and NSF CNS-0960316. HI is supported by JSPS Postdoctoral Fellowships for Research Abroad.  LB was supported by the NSF through grant NSF-DMR-12-06809.

\appendix
\section{Phase Diagram} \label{sec:appA}

In this appendix, we elucidate the ground state phase diagrams for the model in Eq.~\ref{eq:model:Htetra} with interlayer couplings. The phase diagram in absence of interlayer coupling is already presend in Sec.~\ref{sec:independent-layers}, and effect of interlayer couplings for $xy$FM order is discussed in Sec.~\ref{sec:three-dimens-coupl}. Starting from the 2d phase diagram, we here present how the interlayer coupling $K_1$ and $K_2$ modifies the ground state. In Sec.~\ref{ssec:appA:zfm}, we consider $z$FM case. The cases of two AFM orders are considered in Secs.~\ref{ssec:appA:xyafm} and \ref{ssec:appA:zafm}, respectively. In the last, the stripe phase is studied in Sec.~\ref{ssec:appA:stripe}.

\subsection{$z$FM order}\label{ssec:appA:zfm}

For the $z$FM case, a similar phase diagram to the $xy$FM case is obtained. When $K_2<0$, we obtained a simple ferromagnetic order with spins pointing along $z$ axis, as $K_2$ aligns the $z$FM layers ferromagnetically ($z$FM/FM state). On the other hand, when $K_2<0$, the $z$FM planes stacks alternatively forming an antiferromagnetic order ($z$FM/AFM state). These two phases meet at a boundary $K_2=0$.

Introducing $K_1$ gives rise to competition between the $z$FM orders and magnetic states with spins pointing in the $xy$ plane. When $J_1<0$ and $J_1+J_3<0$, the phase diagram consists of four phases: $z$FM/FM, $z$FM/AFM, $xy$FM/FM, and $xy$FM/AFM orders. The two $xy$FM states are separated by the $z$FM states that dominates the phase diagram around $K_1=0$ line. The phase boundaries between the $z$FM and $xy$FM states are given by
\begin{eqnarray}
|K_2^\prime| = |K_1^\prime|+\frac12(J_2-J_1-\frac12J_3).
\end{eqnarray}
The large portion of the phase diagram is covered by the above four phases. When $J_1>0$ and/or $J_1+J_3>0$, however, there remains a small region in which we could not find the ground state by the Luttinger-Tisza method. These regions are given by
\begin{eqnarray}
2J_1|K_2^\prime|^2-J_1(J_2+\frac12J_3) < |K_1^\prime|^2 < J_1^2.
\end{eqnarray}
for $J_3>0$, and
\begin{eqnarray}
2(J_1+J_3)|K_2^\prime|^2-(J_1+J_3)(J_2+\frac12J_3) < |K_1^\prime|^2 < (J_1+J_3)^2.\nonumber\\
\end{eqnarray}
for $J_3<0$. 

\subsection{Antiferromagnetic orders}\label{ssec:appA:xyafm}

The AFM states in the 2d phase diagram in Sec.~\ref{ssec:resA:gs} gives the 3d phase diagram which is quite different from the $xy$FM and $z$FM states. In the case of $xy$AFM state, when $|K_1|$ is small, the energy contribution from $K_1$ bonds cancels out due to the in-plane AFM pattern. Hence, in the classical ground state, arbitrary stacking of the $xy$AFM planes are degenerate as the ground state. The $xy$AFM state with quasi-macriscopic degeneracy is stable as the ground state for 
\begin{eqnarray}
J_1(J_1+J_3) > K_1^2
\end{eqnarray}
and
\begin{eqnarray}
|K_2|<\frac12(J_1+J_2)+\frac14J_3.
\end{eqnarray}
We note that this degeneracy is an accidental one, which will be reduced to two independent sublattices in presence of second neighbor interactions, reducing the ground state to $U(1)\times U(1)$ degrees of freedom. In addition, the $U(1)\times U(1)$ degrees of freedom is expected to be further reduced by quantum fluctuations.

In the classical limit with no further neighbor interactions, this disordered $xy$AFM states are taken over by $xy$FM/FM ($xy$FM/AFM) order for 
\begin{eqnarray}
K_1^\prime{}^2 > \max[(J_1+J_3)^2,J_1^2],
\end{eqnarray}
and $K_1<0$ ($K_1>0$). On the other hand, we could not determine the ground state for the regions in between the $xy$FM and $xy$AFM orders,
\begin{eqnarray}
J_1(J_1+J_3) < K_1^2 < \max[(J_1+J_3)^2,J_1^2].\label{eq:xyafm_incomm}
\end{eqnarray}

Meanwhile, introducing $K_2^\prime$ induce competition between the $xy$ orders and the $z$FM orders. The $xy$AFM orders are stable for
\begin{eqnarray}
|K_2|<\frac12(J_1+J_2)+\frac14J_3,
\end{eqnarray}
and $K_2<0$ ($K_2>0$), and are taken over by the $z$FM/FM ($z$FM/AFM) order for larger $|K_2|$. On the other hand, the $xy$FM/FM and $xy$FM/AFM states that appears in large $|K_1|$ region are stable for
\begin{eqnarray}
|K_2|<|K_1|-\frac12(J_1-J_2+\frac12J_3),
\end{eqnarray}
and taken over by $z$FM/FM ($z$FM/AFM) state when $|K_2|$ is larger and $K_2<0$ ($K_2>0$). The unstable region with moderate $|K_1|$ in Eq.~(\ref{eq:xyafm_incomm}) remains for
\begin{eqnarray}
|K_2|<\frac{K_1^2}{2J_1}-\frac12(J_2-\frac12J_3),
\end{eqnarray}
while the $z$FM/FM ($z$FM/AFM) state takes over for larger $|K_2|$ and $K_2<0$ ($K_2>0$).

\subsection{$z$AFM order} \label{ssec:appA:zafm}

A similar phase diagram to $xy$AFM case is obtained for the $z$AFM case.
The $z$AFM state persists as the ground state in the region
\begin{eqnarray}
(J_1+J_3)(J_2-\frac12J_3)<K_1^2
\end{eqnarray}
and
\begin{eqnarray}
|K_2^\prime| < \frac12(J_1+J_2)+\frac14J_3.
\end{eqnarray}
Similarly to the case of $xy$AFM state, arbitrary stacking of the ordered planes along $z$ axis is allowed, which is expected to be lifted by infinitesimal further neighbor interactions and/or by quantum fluctuation.

With larger $|K_1|$, the $z$AFM phase is taken over by $xy$FM/FM ($xy$FM/AFM) order when $K_1<0$ ($K_1>0$) and
\begin{eqnarray}
K_1^2>\max[(J_1+J_3)^2,J_1^2].
\end{eqnarray}
We could not determine the ground state for the region in between the $z$AFM and $xy$FM/FM ($xy$FM/AFM) order,
\begin{eqnarray}
(J_1+J_3)(J_2-\frac12J_3)<K_1^2<\max[(J_1+J_3)^2,J_1^2]\nonumber\\
\end{eqnarray}
for $J_3>0$ and
\begin{eqnarray}
J_1(J_2+\frac12J_3)<K_1^2<\max[(J_1+J_3)^2,J_1^2]
\end{eqnarray}
for $J_3<0$.

On the other hand, large $|K_2|$ stabilizes $z$FM orders; $z$FM/FM for $K_2<0$ and $z$FM/AFM for $K_2>0$. The phase boundary between $xy$FM and these orders are given by
\begin{eqnarray}
|K_2^\prime| = |K_1^\prime|-\frac12(J_1-J_2-\frac12J_3).
\end{eqnarray}
We also found an unstable region in the phase competing region,
\begin{eqnarray}
K_1^2 > (J_1+J_3)(2|K_2|-J_2-\frac12J_3)
\end{eqnarray}
for $J_3>0$ and
\begin{eqnarray}
K_1^2 > J_1(2|K_2|-J_2+\frac12J_3)
\end{eqnarray}
for $J_3<0$. We could not determine the ground state in these regions.

\subsection{Stripe Orders}\label{ssec:appA:stripe}

In the 2d phase diagram in Sec.~\ref{ssec:resA:gs}, the stripe order appears in the region where $|J_1|$ and $|J_2|$ are relatively small compared to $|J_3|$. In the Luttinger-Tisza method, however, the stripe phase is unstable to infinitesimally small $K_1^\prime$ and $K_2^\prime$. The stripe phase is a fine-tuned case of the ``incommensulate'' ground states. This unstable region appears for 
\begin{eqnarray}
|K_1|&<&J_1+J_3\\
|K_2|&<&\frac{K_1^2}{2J_1} + \frac12(J_2-\frac12J_3)
\end{eqnarray}
when $J_3>0$ and
\begin{eqnarray}
|K_1|&<&J_1\label{eq:stripe_incom}\\
|K_2|&<&\frac{K_1^2}{2J_1} + \frac12(J_2-\frac12J_3)
\end{eqnarray}
if $J_3<0$.
For larger $|K_1|$, the $xy$FM/FM ($xy$FM/AFM) order appears for $K_1<0$ ($K_1>0$) and
\begin{eqnarray}
|K_1|&>&\max(J_1,J_1+J_3).
\end{eqnarray}
On the other hand, $K_2$ stabilize $z$FM/FM ($z$FM/AFM) order for $K_2<0$ ($K_2>0$). It takes over $xy$FM states for
\begin{eqnarray}
|K_2| > |K_1| - \frac12(J_1-J_2+\frac12J_3),
\end{eqnarray}
and the unstable region in Eq.~(\ref{eq:stripe_incom}) for
\begin{eqnarray}
|K_2| > \frac{K_1^2}{2J_1} + \frac12(J_2-\frac12J_3).
\end{eqnarray}

\end{document}